\documentclass[pdflatex,sn-mathphys-num]{sn-jnl}
\usepackage{xcolor}
\usepackage{booktabs}
\usepackage{graphicx}
\usepackage{pgfplots}
\usepackage{ifthen}
\usepackage{subcaption}
\usepackage{import}

\usepgfplotslibrary{colormaps}
\usepgfplotslibrary{statistics}
\pgfplotsset{compat=newest} 
\pgfdeclarelayer{background}
\pgfdeclarelayer{background_1}
\pgfdeclarelayer{background_2}
\pgfdeclarelayer{background_3}
\pgfdeclarelayer{foreground_6}
\pgfdeclarelayer{foreground_5}
\pgfdeclarelayer{foreground_4}
\pgfdeclarelayer{foreground_3}
\pgfdeclarelayer{foreground_2}
\pgfdeclarelayer{foreground_1}
\pgfdeclarelayer{foreground}
\pgfsetlayers{background,background_1,background_2,background_3,main,foreground_6,foreground_5,foreground_4,foreground_3,foreground_2,foreground_1,foreground}

\usepackage{tikz}
\usetikzlibrary{arrows, arrows.meta, positioning, quotes, shapes, fit, calc, decorations.pathreplacing, calligraphy}
\tikzstyle{block} = [draw, thick, text=black, align=center, minimum width=1.2cm, minimum height=0.6cm, font=\small]
\tikzstyle{Block} = [draw, thick, rounded corners, text=black, align=center, minimum width=2cm, minimum height=1.2cm, font=\small]
\usepackage{circuitikz}
\ctikzset{bipoles/cuteswitch/thickness=0.3}
\usetikzlibrary{external,plotmarks,fit, fadings}

\usepackage{amsmath}
\usepackage{siunitx}
\usepackage{mleftright}
\usepackage{relsize}
\usepackage{svg}
\usepackage{svg-extract}

\usepackage{textgreek}
\usepackage{mathtools}
\usepackage{float} 

\usepackage[nohyperlinks,nolist]{acronym}

\acrodef{1dconv}[1D-Conv]{one-dimensional convolution}

\acrodef{ai}[AI]{artificial intelligence}
\acrodef{ask}[ASK]{amplitude-shift keying}
\acrodef{awgn}[AWGN]{additive white Gaussian noise}
\acrodef{ann}[ANN]{artificial neural network}
\acrodef{asic}[ASIC]{application-specific integrated circuit}

\acrodef{bce}[BCE]{binary cross-entropy}
\acrodef{ber}[BER]{bit error ratio}
\acrodef{bler}[BLER]{block error ratio}
\acrodef{bpsk}[BPSK]{binary phase-shift keying}
\acrodef{bram}[BRAM]{block random access memory}
\acrodefplural{bram}[BRAMs]{block random access memories}
\acrodef{bilstm}[biLSTM]{bidirectional long short-term memory}
\acrodef{bp}[BP]{backward pass}

\acrodef{cd}[CD]{chromatic dispersion}
\acrodef{cir}[CIR]{combined impulse response}
\acrodef{cma}[CMA]{constant modulus algorithm}
\acrodef{cnn}[CNN]{convolutional neural network}
\acrodef{cpu}[CPU]{central processing unit}
\acrodef{ctp}[CTP]{channel transition probability}
\acrodefplural{ctp}[CTPs]{channel transition probabilities}
\acrodef{csi}[CSI]{channel state information}

\acrodef{dnn}[DNN]{deep neural network}
\acrodef{dsp}[DSP]{digital signal processor}
\acrodef{dram}[DRAM]{dynamic random access memory}
\acrodef{dfe}[DFE]{decision-feedback equalizer}
\acrodef{dop}[DOP]{degree of parallelism}

\acrodef{fc}[FC]{fully connected}
\acrodef{fec}[FEC]{forward error correction}
\acrodef{fir}[FIR]{finite impulse response}
\acrodef{fifo}[FIFO]{first in first out}
\acrodef{fpga}[FPGA]{field programmable gate array}
\acrodef{fp}[FP]{forward pass}
\acrodef{ff}[FF]{flip-flop}

\acrodef{elu}[ELU]{exponential linear unit}

\acrodef{gan}[GAN]{generative adversarial network}
\acrodef{gpu}[GPU]{graphics processing unit}

\acrodef{hls}[HLS]{high-level synthesis}
\acrodef{hwa}[HWA]{historical weight averaging}
\acrodef{ht}[HT]{high-throughput}

\acrodef{imdd}[IM/DD]{intensity modulation with direct detection}
\acrodef{iot}[IoT]{Internet of things}
\acrodef{isi}[ISI]{inter-symbol interference}

\acrodef{lms}[LMS]{least-mean-square}
\acrodef{lut}[LUT]{look-up table}
\acrodef{lp}[LP]{low-power}

\acrodef{mlsd}[MLSD]{maximum-likelihood sequence detection}
\acrodef{mse}[MSE]{mean-squared-error}
\acrodef{mac}[MAC]{multiply-accumulate}
\acrodef{ma}[MA]{moving average}
\acrodef{map}[MAP]{maximum a posteriori}
\acrodef{msm}[MSM]{merge stream module}
\acrodef{mzm}[MZM]{mach-zehnder modulator}
\acrodef{nn}[NN]{neural network}
\acrodef{ns}[NS]{non-saturating}

\acrodef{ogm}[OGM]{overlap generate module}
\acrodef{orm}[ORM]{overlap remove module}

\acrodef{pam}[PAM]{pulse-amplitude modulation}
\acrodef{pdf}[pdf]{probability density function}
\acrodef{pe}[PE]{processing element}
\acrodef{pmf}[pmf]{probability mass function}

\acrodef{qlf}[QLF]{quantization trade-off factor}
\acrodef{rc}[RC]{raised-cosine}
\acrodef{relu}[ReLU]{rectified linear unit}
\acrodef{rls}[RLS]{recursive least squares}
\acrodef{rnn}[RNN]{recurrent neural network}

\acrodef{ser}[SER]{symbol error rate}
\acrodef{sgd}[SGD]{stochastic gradient descent}
\acrodef{sld}[SLD]{square-law detection}
\acrodef{snr}[SNR]{signal-to-noise ratio}
\acrodef{sps}[sps]{samples per symbol}
\acrodef{ssmf}[SSMF]{standard single-mode fiber}
\acrodef{simd}[SIMD]{single instruction multiple data}
\acrodef{ssm}[SSM]{split stream module}
\acrodef{spb}[SPB]{symbols per batch}

\acrodef{ti}[TI]{training iteration}

\acrodef{vnle}[VNLE]{Volterra-based nonlinear equalizer}

\DeclarePairedDelimiter\floor{\lfloor}{\rfloor}

\newcommand{\mlrb}[1]{\mleft\{#1\mright\}}

\definecolor{RPTU_BlueGray}{RGB}{80,114,137}
\definecolor{RPTU_GreenGray}{RGB}{119,182,186}
\definecolor{RPTU_DarkBlue}{RGB}{4,44,88}
\definecolor{RPTU_LightBlue}{RGB}{106,178,231}
\definecolor{RPTU_DarkGreen}{RGB}{0,107,107}
\definecolor{RPTU_LightGreen}{RGB}{38,208,124}
\definecolor{RPTU_Violett}{RGB}{76,53,117}
\definecolor{RPTU_Pink}{RGB}{209,56,150}
\definecolor{RPTU_Red}{RGB}{227,27,76}
\definecolor{RPTU_Orange}{RGB}{255,162,82}
\definecolor{RPTU_Black}{RGB}{0,0,0}
\definecolor{RPTU_White}{RGB}{255,255,255}

\definecolor{customOrange}{RGB}{255,127,0}
\definecolor{customBlue}{RGB}{55,126,184}
\definecolor{limeGreen}{RGB}{50,205,50}  
\definecolor{turquoise}{RGB}{64, 224, 208}
\definecolor{customRed}{RGB}{228,26,28}   
\definecolor{customYellow}{RGB}{210,210,0}
\definecolor{customGreen}{RGB}{77,175,74}

\newcommand{\new}[1]{{\leavevmode\color{black}#1}}

\begin{document}

\title{CNN-Based Equalization for Communications: Achieving Gigabit Throughput with a Flexible FPGA Hardware Architecture}

\author*[1]{\fnm{Jonas} \sur{Ney}}\email{jonas.ney@rptu.de}

\author[2]{\fnm{Christoph} \sur{Füllner}}\email{christoph.fuellner@kit.edu}

\author[3]{\fnm{Vincent} \sur{Lauinger}}\email{vincent.lauinger@kit.edu}

\author[3]{\fnm{Laurent} \sur{Schmalen}}\email{laurent.schmalen@kit.edu}

\author[2]{\fnm{Sebastian} \sur{Randel}}\email{sebastian.randel@kit.edu}

\author[1]{\fnm{Norbert} \sur{Wehn}}\email{norbert.wehn@rptu.de}

\affil*[1]{\orgdiv{Microelectronic Systems Design (EMS)}, \orgname{RPTU Kaiserslautern-Landau}, \orgaddress{\city{Kaiserslautern}, \postcode{67653}, \state{Germany}}}

\affil[2]{\orgdiv{Institute of Photonics and Quantum Electronics (IPQ)}, \orgname{KIT}, \orgaddress{\city{Karlsruhe}, \postcode{76131}, \state{Germany}}}

\affil[3]{\orgdiv{Communications Engineering Lab (CEL)}, \orgname{KIT}, \orgaddress{\city{Karlsruhe}, \postcode{76131}, \state{Germany}}}

\abstract{
To satisfy the growing throughput demand of data-intensive applications, the performance of optical communication systems increased dramatically in recent years. With higher throughput, more advanced equalizers are crucial, to compensate for impairments caused by \ac{isi}. The latest research shows that \ac{ann}-based equalizers are promising candidates to replace traditional algorithms for high-throughput communications. On the other hand, not only throughput but also flexibility is a main objective of beyond-5G and 6G communication systems. A platform that is able to satisfy the strict throughput and flexibility requirements of modern communication systems are \acp{fpga}. Thus, in this work, we present a high-performance \ac{fpga} implementation of an \ac{ann}-based equalizer, which meets the throughput requirements of modern optical communication systems. Further, our architecture is highly flexible since it includes a variable \ac{dop} and therefore can also be applied to low-cost or low-power applications which is demonstrated for a magnetic recording channel. The implementation is based on a cross-layer design approach featuring optimizations from the algorithm down to the hardware architecture, including a detailed quantization analysis. Moreover, we present a framework to reduce the latency of the \ac{ann}-based equalizer under given throughput constraints. As a result, the \ac{ber} of our equalizer for the optical fiber channel is around four times lower than that of a conventional one, while the corresponding \ac{fpga} implementation achieves a throughput of more than \SI{40}{\giga Bd}, outperforming a high-performance \ac{gpu} by three orders of magnitude for a similar batch size. 
}


\maketitle


\acresetall
\section{Introduction}

In recent years, the achievable throughput of optical communication systems grew dramatically, driven by the increasing demand for high-speed data transmission in various applications such as data centers, video streaming, and cloud computing~\cite{Bayvel2016}. The higher throughput leads to lower \ac{snr} and an increased \ac{isi} which strongly impairs the system's performance. As a result, the design and implementation of advanced signal processing techniques has become crucial for maintaining high communication performance and low \ac{ber}. 
On the other hand, in addition to high throughput, future communication standards strongly focus on flexible architectures~\cite{Yazar2020},~\cite{Viswanathan2020} to satisfy application requirements for various use cases. 

To prevent an increase in \ac{ber} while providing high flexibility, latest research put a strong emphasis on \ac{ann}-based algorithms for communication systems~\cite{zerguine2001},~\cite{schaedler2019},~\cite{ney2022}. Especially the equalizer, responsible for compensating channel impairments of the received signal, is a component that can benefit from the advancements of \ac{ann} research. In particular, \acp{ann} have shown remarkable results for channels with non-linear effects, for which no exact analytical solutions exist for equalization \cite{Khan2019}. In this work, we study a  \SI{40}{\giga Bd} optical \ac{imdd} channel, where non-linear distortions are caused by \ac{cd} \cite{Amari2017}, as an exemplary use case for a high-throughput \ac{ann}-based equalization. Further, we provide results for a low-cost telephone channel to highlight the flexibility of our hardware architecture. 

Compared to conventional algorithms, which have been optimized for decades, \acp{ann} introduce high computational complexity, limiting the achievable throughput on general-purpose processors like \acp{cpu} or \acp{gpu}. In contrast, application-specific platforms like \acp{fpga} or \acp{asic} provide huge parallelism and customizability. As compared to \acp{asic}, the deployment of \acp{fpga} is more cost-efficient for low-volume productions and they can provide a shorter time-to-market \cite[Ch.~1]{kuon2010quantifying}. Further, \acp{fpga} can be reprogrammed to provide the required flexibility. 

Most previous works utilized \acp{fpga} as a platform for performance evaluation and prototyping of \ac{asic} implementations \cite{Freire2022},~\cite{Kaneda2022},~\cite{Li2021},~\cite{Huang2022},~\cite{Ney2023}. In contrast, in this work, we aim to satisfy the strict throughput requirements on the \ac{fpga} itself. This way the design can either be deployed on the \ac{fpga} or be used as a well-grounded base for an \ac{asic} design. However, when the design is deployed on an \ac{fpga}, it has to satisfy the strict performance requirements on this resource-constrained device, which is a great challenge even with advanced \ac{fpga} platforms.
In~\cite{Freire2022}, the \ac{fpga} implementation of a \ac{rnn}-based equalizer for a \SI{34}{\giga Bd} single-channel dual-polarization as well as approximations of non-linear activation functions are presented. Their throughput requirements can be satisfied by using \num{5} \acp{fpga} and a high number of parallel outputs (\num{61}) of each \ac{rnn}, which is not feasible for the \SI{40}{\giga Bd} \ac{imdd} channel we focus on in this work, as it results in significantly increased \ac{ber}. 
In \cite{Kaneda2022} and \cite{Li2021}, a channel more similar to ours is considered. However, the implementation of \cite{Kaneda2022} only achieves a throughput of \SI{2.6}{Gbps} for a channel with a data-rate of \SI{50}{Gbps}. In \cite{Li2021} a pruned \ac{ann} is utilized for equalization of a \SI{50}{\giga Bd} \ac{pam}-4 channel. To achieve the required throughput on \ac{fpga}, the \ac{ber} requirement needs to be relaxed from \num{1e-5} to \num{3.8e-3}. 
In \cite{Huang2022}, an \ac{rnn} is compared to a fully-connected \ac{ann} for equalization of an \ac{imdd} channel. While the \ac{rnn} achieves a much lower \ac{ber}, only the \ac{ann} is able to meet the throughput requirements of the optical communication channel. 
In \cite{Ney2023} a novel unsupervised loss function for ANN-based equalization is proposed. Further, a trainable \ac{fpga} implementation of the approach is presented which enables adaptation to varying channel conditions during runtime. However, the implementation does not achieve the required channel throughput of \SI{25}{\giga Bd}.

In this work, we present a high-throughput \ac{fpga} implementation of a \ac{cnn}-based equalizer for an optical \ac{imdd} channel. In contrast to previous works, we apply a cross-layer design methodology, which involves an extensive design space exploration of the \ac{cnn} topology, and a framework for selecting the appropriate sequence length per \ac{cnn} instance. Moreover, a detailed quantization analysis based on an automatic quantization approach is conducted. Further, we show how our design approach can also be applied to low-cost channels with lower throughput constraints.
For the high-throughput, \SI{40}{\giga Bd} channel, we focus on high parallelism across all design layers, from algorithm down to implementation. Further, special attention is given to low latency which is crucial for optical communication used in high-frequency trading or telemedicine.

As a result, our \ac{fpga} implementation achieves a \ac{ber} around one order of magnitude lower than that of a conventional equalizer, while satisfying the throughput requirement of \SI{40}{\giga Bd}. Further, our approach outperforms an implementation on a high-performance 
\ac{gpu} by four orders of magnitude for a similar batch size. 
\newpage
This article is an extension of the work presented in \cite{Ney2023_2}. It significantly extends the previous publication by introducing \ac{fir} filters and Volterra kernels to the design space exploration to provide a fairer comparison, by conducting a comprehensive quantization analysis, by extending the hardware architecture with a flexible \ac{dop} to enable the adaptation to different application scenarios, by applying and evaluating the approach for a magnetic recording channel, by analyzing the influence of the \ac{dop} on the throughput and the power consumption, and by extending the comparisons to TensorRT implementations, embedded GPU implementations and providing an analysis of the dynamic power consumption. 
\hfill\\
\hfill\\
\noindent
In summary, our novel contributions are: 
\begin{itemize}
    \item[$\bullet$] A detailed design space exploration of the \ac{cnn}, featuring cross-layer analysis and automatic quantization, resulting in a network with a \ac{ber} one order of magnitude lower than that of a conventional equalizer;
    \item[$\bullet$] An efficient hardware architecture, suited for high-throughput as well as low-cost application scenarios by providing flexibile \acp{dop} on multiple implementation levels
    \item[$\bullet$] An in-depth trade-of analysis of our automatic quantization approach 
    \item[$\bullet$] A framework allowing to trade-off throughput against latency to adapt for application requirements, based on a timing model of our architecture;
    \item[$\bullet$] An advanced implementation of a high-performance \ac{cnn}-based equalizer for optical communication, achieving a throughput of more than \SI{40}{\giga Bd}
\end{itemize}

\section{Investigated Communication Channels}
\label{sec:communication_channel}
For our \ac{cnn}-based equalization approach we mainly focus on the high-throughput \ac{imdd} channel presented in Sec. \ref{sec:experimental_setup}. To further highlight the flexibility of our approach we also give results for the band-limited magnetic recording channel as described in Sec. \ref{sec:telephone_channel}. The first channel is based on an experimental setup where the input and output data is captured while the second channel is simulated in a Python environment. 

\subsection{Fiber-Optical Channel}
\label{sec:experimental_setup}

In fiber-optic communications, the application of \ac{ann}-based equalizers has been proven beneficial for mitigation of nonlinear distortions for which no analytic expression exists \cite{Owaki2016, Estaran2016}.
In this work, as an example of a high-throughput channel, we chose an \ac{imdd} transmission system, where non-linear impairments are caused by the interplay between \ac{cd} and direct detection to evaluate the performance of the \ac{ann}-based equalizer.
Specifically, we modulate the intensity of a continuous-wave laser tone at \SI{1550}{\nano \meter} with a high-speed zero-chirp \ac{mzm} that is biased at the quadrature point. Following the recommendations of \cite{Eriksson2017}, we use a pseudo random sequence based on the Mersenne-Twister algorithm as a transmit pattern and drive the \ac{mzm} with a \SI{40}{\giga Bd} pulse amplitude modulation signal with two levels (PAM2) and a root-raised-cosine spectral shape. 
The resulting optical on-off-keying signal is launched into a standard single-mode fiber with a length of \SI{31.5}{\kilo \meter} that features a \ac{cd} coefficient of approximately \SI[inter-unit-product =\,]{16}{\pico\second\per\nano\meter\per\kilo\meter}. At the receiver side, we employ a \SI{40}{\giga \hertz} photodetector to detect the envelope of the optical signal. Since \ac{cd} is an effect related to the optical field, it impairs the photocurrent obtained after square-law detection in a nonlinear way. 
Finally, the electrical signal is recorded by a real-time oscilloscope. We digitally resample the captured waveforms and apply a timing recovery algorithm to align the received waveform with the transmit pattern for the training of the \ac{ann}. 
We digitally precompensate the frequency-dependent attenuation of the transmitter components so that transceiver noise and \ac{cd} remain as the effects impairing the quality of the received signal.

\new{
\subsection{Magnetic Recording Channel}
\label{sec:telephone_channel}
As a second channel, with a smaller bandwidth and therefore a lower maximal throughput, we simulate a linear bad-quality communication channel as described in \cite[Ch.~9.4-3]{proakis2008digital}. The channel is known as \textit{Proakis-B} and has the following discrete impulse response: 
\begin{align*}
	\mathbf{h}_{\mathrm{ch,\, ProB}} &= [0.407, \ 0.815, \ 0.407] \; .
\end{align*}

In the simulation, the transmitted symbols $\mathbf{x}$, are convolved with a \ac{rc} pulse shaping filter and the linear channel impulse response of the channel $h_{\mathrm{ch,\, ProB}}$.
Afterwards, the received vector is superimposed by Gaussian noise. Similar to the experimental setup, we run our simulation with an oversampling rate of $N_{\mathrm{os}}=2$.
}
\section{CNN Design Space Exploration}
\label{sec:dse}

The following design space exploration is performed for the optical fiber channel described in Sec. \ref{sec:experimental_setup}. At the end of the section, we show that the obtained model can also be successfully applied to the simulated magnetic recording channel. 

An optimized neural network topology is crucial for hardware implementation of \ac{ann}-based algorithms, as it has a huge influence on the power consumption, throughput, and latency of the final implementation. However, the design of efficient \ac{ann} topologies is characterized by an enormous design space with various hyperparameters. This design space includes the layer type, the number of layers, the size of each layer, the activation function, and multiple more hyperparameters. An exploration of all those parameters is nearly infeasible, thus we restrict our analysis to the topology template presented in the following, which provides sufficient configurability while comprising a manageable design space.  

\subsection{CNN Topology Template}
\label{sec:topology}

\begin{figure}[t!]
    \vspace{-4mm}
	\centerline{\tikzsetnextfilename{CNN_topology}
\begin{tikzpicture}[node distance=0.2,>=latex]
    \tikzset{near start abs/.style={xshift=.01cm}}
    \def\minWid{3cm}; \def\minHei{0.7cm}; \def\arrowLen{0.7cm};
    \def\boxFontSize{\scriptsize}

    \node (G_in) {};
    \node[block, rounded corners, draw=RPTU_DarkGreen, minimum width=\minWid, minimum height=\minHei, below=\arrowLen*0.8 of G_in] (G_conv_1) {Conv1D: $K$, $S$=$V_p$};
    \node[block, rounded corners, draw=RPTU_Violett, minimum width=\minWid, minimum height=\minHei, below=0.04cm of G_conv_1] (G_batchnorm_1) {Batchnorm, ReLU};
    
    \node[right=-0.1cm of G_batchnorm_1] (dot_begin) {};
    \node[right=0.45*\arrowLen of dot_begin] (dot_end) {};

    \node[block, rounded corners, draw=RPTU_DarkGreen, minimum width=\minWid, minimum height=\minHei, right=\arrowLen*1.5 + 0.5cm of G_conv_1] (G_conv_2) {Conv1D: $K$, $S$=$1$};
    \node[block, rounded corners, draw=RPTU_Violett, minimum width=\minWid, minimum height=\minHei, below=0.04cm of G_conv_2] (G_batchnorm_2) {Batchnorm, ReLU};

    \node[block, rounded corners, draw=RPTU_DarkGreen, minimum width=\minWid, minimum height=\minHei, right=\arrowLen*2 of G_batchnorm_2, yshift=0.25cm] (G_conv_3) {Conv1D: $K$, $S$=$N_\textrm{os}$};
    
    \node[below=\arrowLen*0.8 of G_conv_3] (G_out) {};

    \node[below=0.05cm of G_batchnorm_1] (l1) {\normalsize $1$};
    \node[below=0.05cm of G_batchnorm_2] (l_L-2) {\normalsize $L-1$};
    \node[below=0.3cm of G_conv_3, xshift=-0.6cm] (l_L-1) {\normalsize $L$};

    \draw[-{Latex[length=2mm]}, thick] (G_in) -- node[midway, right, xshift=0.1cm, yshift=0.05cm] {$1 \times S_\mathrm{in}$} (G_conv_1.north);

    \draw[dotted, thick] ($(dot_begin.east) + (-0.1cm, +0.25cm)$) -- ($(dot_end.west) + (0.05cm, +0.25cm)$);

    \draw[-{Latex[length=2mm]}, thick] ($(dot_end.east) + (-0.1cm, +0.25cm)$) -- node[midway, above, xshift=-0.225cm, yshift=0.05cm] {$C \times \frac{S_\mathrm{in}}{V_p}$} ($(G_batchnorm_2.west) + (-0.1cm, +0.25cm)$);
    \draw[-{Latex[length=2mm]}, thick] ($(G_batchnorm_2.east) + (0.1cm, +0.25cm)$) -- node[midway, above, xshift=0cm, yshift=0.05cm] {$C \times \frac{S_\mathrm{in}}{V_p}$} ($(G_conv_3.west) + (-0.1cm, 0cm)$);
    \draw[-{Latex[length=2mm]}, thick] (G_conv_3) -- node[midway, right, xshift=0.1cm, yshift=-0.1cm] {$V_p \times \frac{S_\mathrm{in}}{V_p \cdot N_\textrm{os}}$} (G_out);

\end{tikzpicture}}
	\caption{Topology template of the equalizer \ac{cnn}. The feature map dimensions are given next to the arrows, where the first dimension corresponds to the number of channels and the second one to the width.}
 	\label{fig:network_topology_template}
\end{figure}
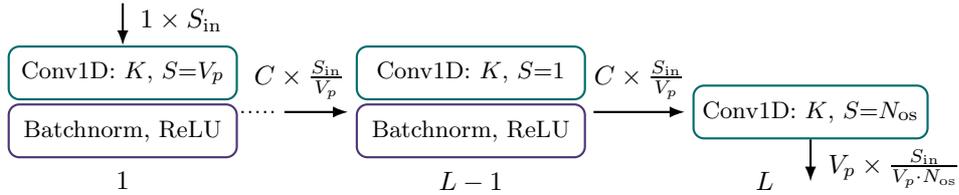

The core of our equalizer is a one-dimensional \ac{cnn} since it resembles the structure of traditional convolutional filters. The \ac{cnn} is based on a customizable topology template shown in Fig. \ref{fig:network_topology_template}, where specific parameters are determined in an extensive design space exploration. 

The \ac{cnn} is composed of $L$ convolutional layers with identical kernel size $K$ and padding $P$. Each convolutional layer but the last is followed by batch normalization and \ac{relu} activation functions. One channel is used for the input sequence, while subsequent activations consist of $C$ channels. The output of the \ac{cnn} is based on $V_\mathrm{p}$ channels, thus $V_\mathrm{p}$ values are calculated in parallel for one pass of the network. To shift the input sequence accordingly, the first layer has a stride of $V_\mathrm{p}$, while the following layers have a stride of one, and the last stride is set corresponding to the oversampling factor $N_\mathrm{os}$. 
After the last convolutional layer, the feature map is flattened so that each element of the feature map corresponds to one output symbol. Afterwards, the output is mapped to the closest constellation symbol.

\new{

\subsection{Linear Equalizer}

To compare the performance of our \ac{cnn}-based equalizer, we also include a conventional linear feedforward equalizer in our design space exploration. The equalizer is based on a \ac{fir} filter with $M$ taps, which is convolved with the input sequence to calculate the output sequence. For a time-discrete system, the output sequence is the weighted sum of $M$ inputs:

\begin{equation}
\label{eq:fir}
    y_{i} = \sum_{m=-M^*}^{M^*} x_{i+m} \cdot w(m + M^*)\; , 
\end{equation}

where $x$ is the input sequence, $y$ is the output sequence, $w(m)$ is the weight of the $m\mathrm{th}$ tap,  $M^* = \floor*{M \mathbin{/} 2}$ and $\floor*{\cdot}$ declares the floor operator. For an oversampling factor of $N_\mathrm{os}=2$, every second output sample corresponds to an output symbol. Similar to the \ac{cnn}, after equalization, the output is mapped to the closest constellation symbol. We train the linear equalizer in a supervised manner using the \ac{mse} loss and the Adam optimizer.  

\subsection{Volterra Equalizer}
For further comparison, we also include a more complex equalizer which is based on the Volterra kernel. This nonlinear equalizer calculates the sum of multidimensional convolutions with Volterra kernels up to an order of $P$. Since high-order Volterra kernels are associated with high complexity, we restrict our exploration up to kernels of order \num{3}. A Volterrra equalizer of order \num{3} can be described by the following equation: 
\begin{align*}
    &y_{i} ={} w_0 + \sum_{m_1=-M_1^*}^{M_1^*} x_{i + m_1} \cdot w_1(m_1 + M_1^*) +\\ &\sum_{m_1=-M_2^*}^{M_2^*}\sum_{m_2=-M_2^*}^{M_2^*} x_{i + m_1} \cdot x_{i + m_2}  \cdot w_2(m_1 + M_2^*, m_2 + M_2^*) +\\ &\sum_{m_1=-M_3^*}^{M_3^*}\sum_{m_2=-M_3^*}^{M_3^*}\sum_{m_3=-M_3^*}^{M_3^*}  x_{i + m_1} \cdot x_{i + m_2} \cdot x_{i + m_3} \cdot w_3(m_1 + M_3^*, m_2 + M_3^*, m_3 + M_3^*) \; , 
\end{align*}
where $x$ is the input sequence, $y$ is the output sequence, $M_p$ is the memory length of the $p\mathrm{th}$ order kernel,  $M_p^* = \floor*{M_p \mathbin{/} 2}$, $\floor*{\cdot}$ declares the floor operator, $w_1$ are the first-order weights, $w_2$ are the second-order weights and $w_3$ are the third-order weights. Similar to the \ac{cnn}-based and the linear equalizer, the Volterra equalizer is trained in a supervised fashion with \ac{mse} loss and Adam optimizer. 

}

\subsection{Design Space Exploration Framework}

To explore this design space of various \ac{cnn} configurations, we design a framework that allows us to automatically evaluate multiple configurations which are compared in terms of communication performance and complexity. Further, the framework features cross-layer analysis by providing an estimate of the achievable throughput. Thus hardware metrics are already included in the topology search, which greatly reduces the development cycles since multiple models can already be discarded in an early design phase. 

As configurable hyperparameters of the \ac{cnn}, we select the number of layers $L$, the kernel size $K$, the number of channels $C$, and the symbols calculated in 
parallel $V_p$. We train each configuration three times for \num{10000} iterations with an initial learning rate of \num{0.001} with the Adam optimizer and the \ac{mse} loss. 
After training, the highest achieved \ac{ber} of the three training runs and the corresponding \ac{mac} operations per input symbol of each configuration are determined by our framework. This way, a trade-off between communication performance and hardware complexity can be found. 

\new{For the linear equalizer and the Volterra equalizer, we explore the design space by varying the number of taps. While this corresponds to a one-dimensional exploration for the linear equalizer, the Volterra equalizer contains taps in three dimensions. Similar to the \ac{cnn}, we evaluate the complexity of both equalizers based on the number of \ac{mac} operations to calculate one output symbol. 
}

\subsection{Results of Design Space Exploration}

In Fig. \ref{fig:dse}, the results of the design space exploration are shown. Our design space for the \ac{cnn} is spanned by the following four dimensions: symbols calculated in parallel  $V_p \in \mlrb{1, 2, 4, 8, 16}$, network depth $L \in \mlrb{3, 4, 5}$, kernel size $K \in \mlrb{9, 15, 21}$ and the number of channels $C \in \mlrb{3, 4, 5}$. Thus, overall \num{135} different models are trained and evaluated. The average \ac{mac} operations per symbol $\mathrm{MAC}_\mathrm{sym}$ can be calculated as follows: 

\begin{equation*}
    \mathrm{MAC}_\mathrm{sym} = \frac{K \cdot C}{V_p} + (L - 2) \cdot \frac{K \cdot C \cdot C}{V_p} + \frac{K \cdot C}{N_\mathrm{os}} \; .
\end{equation*}

\new{For the linear equalizer, the design space is spanned by the number of taps $M \in \mlrb{3, 5, 9, 17, 25, 41, 57, 89, 121, 185, 249, 377, 505, 761, 1017}$ and for the Volterra equalizer by the number of taps of each order $M_1 \in \mlrb{3, 9, 15, 25, 35, 55, 75, 89, 121}$, $M_2 \in \mlrb{1, 3, 9, 15, 25, 30, 35}$ and $M_3\in \mlrb{1, 3, 9, 15}$.

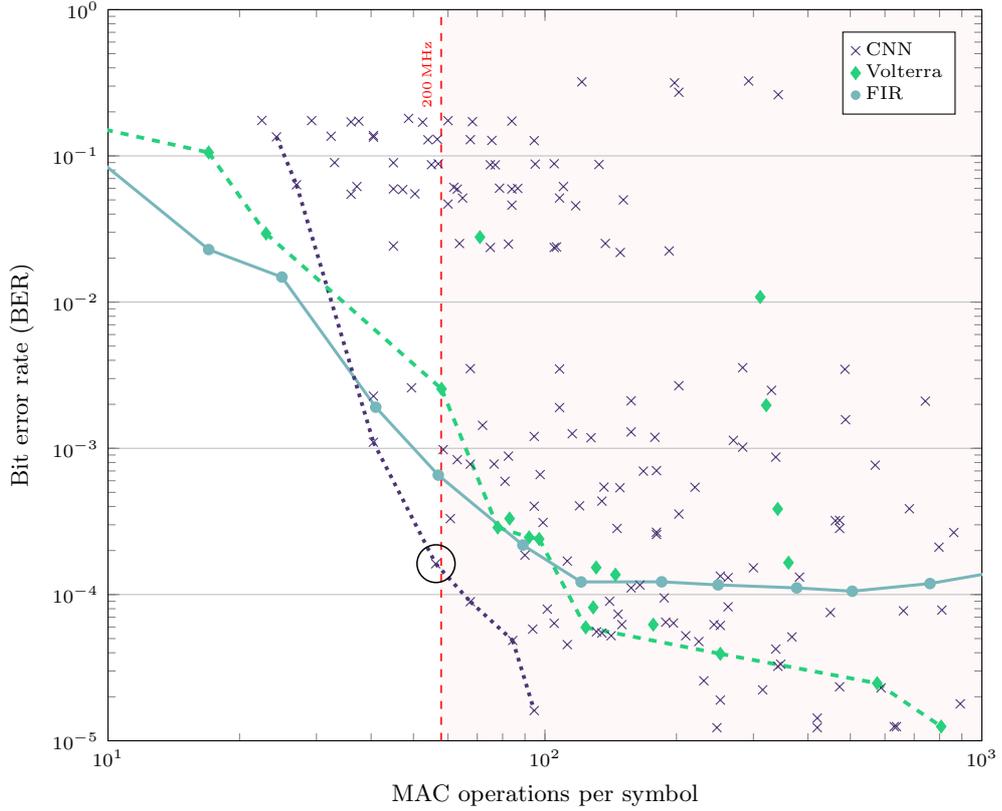
\begin{figure}[b!]
	\centering
	\newboolean{do_plot_linear}   
\newboolean{do_plot_volterra}   
\newboolean{do_plot_pareto_cnn}   
\newboolean{do_plot_pareto_volterra}   
\newboolean{do_plot_compl_bound}   
\newboolean{do_plot_sel_model}

\setboolean{do_plot_pareto_cnn}{true} 
\setboolean{do_plot_pareto_volterra}{true} 
\setboolean{do_plot_linear}{true} 
\setboolean{do_plot_volterra}{true} 
\setboolean{do_plot_compl_bound}{true} 
\setboolean{do_plot_sel_model}{true}

\definecolor{customOrange}{RGB}{255,127,0}
\definecolor{customBlue}{RGB}{55,126,184}
\definecolor{limeGreen}{RGB}{50,205,50}  
\definecolor{turquoise}{RGB}{64, 224, 208}
\definecolor{customRed}{RGB}{228,26,28}   
\definecolor{customYellow}{RGB}{210,210,0}
\definecolor{customGreen}{RGB}{77,175,74} 

\tikzsetnextfilename{DSE}

\begin{tikzpicture}
    [
    square/.style={regular polygon,regular polygon sides=4},
	/pgfplots/every mark/.append style={solid, mark size=3.5pt},
	/pgfplots/every axis/.append style={font=\small}, 
	/pgfplots/every tick label/.append style={font=\footnotesize},
	/pgfplots/every axis legend/.append style={nodes={right},font={\footnotesize},row sep=-2pt},
	]

	\begin{axis}[
             xmin=10,
            xmax=1000,
            ymin=1e-5,
            ymax=1,
            xmode=log,
            ymode=log, 
            legend pos=north east,
            xlabel={MAC operations per symbol},
            ylabel={Bit error rate (BER)},
            width=1*\columnwidth,
	        ymajorgrids
            ]

    \addplot [only marks, mark=x, color=RPTU_Violett, mark options={scale=1.2}] table [x=MAC,y=BER]{data/DSE/all_points_no_bias.txt}; 
    \addlegendentry{CNN};

    \ifthenelse{\boolean{do_plot_volterra}}{
        \addplot [only marks, mark=diamond*, color=RPTU_LightGreen, mark options={scale=1.2, fill=RPTU_LightGreen}] table [x=MAC,y=BER]{data/DSE/volterra_new.txt}; 
        \addlegendentry{Volterra};
    }{}
    
    \ifthenelse{\boolean{do_plot_linear}}{
        \addlegendimage{only marks, mark=*, mark options={scale=0.8}, color=RPTU_GreenGray};
        \addlegendentry{FIR};
        \addplot [solid, mark=*, color=RPTU_GreenGray, mark options={scale=0.8}, line width=0.4mm] table [x=MAC,y=BER]{data/DSE/fir_filter.txt}; 
    }{}

    \ifthenelse{\boolean{do_plot_pareto_cnn}}{
        \draw [dotted, RPTU_Violett, line width=0.5mm] (24.3, 0.13485535) --  (27, 0.063423939);
        \draw [dotted, RPTU_Violett, line width=0.5mm] (27, 0.063423939) --  (40.5, 0.001100779);
        \draw [dotted, RPTU_Violett, line width=0.5mm] (40.5, 0.001100779) --  (56.25, 0.000162292);
        \draw [dotted, RPTU_Violett, line width=0.5mm] (56.25, 0.000162292) --  (67.5, 8.9434E-05);
        \draw [dotted, RPTU_Violett, line width=0.5mm] (67.5, 8.9434E-05) --  (84.375, 4.85285E-05);
        \draw [dotted, RPTU_Violett, line width=0.5mm] (84.375, 4.85285E-05) --  (94.5, 1.60908E-05);
    }{}

    \ifthenelse{\boolean{do_plot_pareto_volterra}}{
        \draw [dashed, RPTU_LightGreen, line width=0.5mm] (6, 0.20953046) --  (17, 0.10549016);
        \draw [dashed, RPTU_LightGreen, line width=0.5mm] (17, 0.10549016) --  (23, 0.02944212);
        \draw [dashed, RPTU_LightGreen, line width=0.5mm] (23, 0.02944212) --  (58, 0.002548);
        \draw [dashed, RPTU_LightGreen, line width=0.5mm] (58, 0.002548) --  (78, 0.0002873);
        \draw [dashed, RPTU_LightGreen, line width=0.5mm] (78, 0.0002873) --  (92, 0.0002462);
        \draw [dashed, RPTU_LightGreen, line width=0.5mm] (92, 0.0002462) --  (97, 0.0002399);
        \draw [dashed, RPTU_LightGreen, line width=0.5mm] (97, 0.0002399) --  (124, 5.9609e-05);
        \draw [dashed, RPTU_LightGreen, line width=0.5mm] (124, 5.9609e-05) --  (252, 3.93419e-05);
        \draw [dashed, RPTU_LightGreen, line width=0.5mm] (252, 3.93419e-05) --  (576, 2.473772e-05);
        \draw [dashed, RPTU_LightGreen, line width=0.5mm] (576, 2.473772e-05) --  (807, 1.251788e-05);
    }{}

    \ifthenelse{\boolean{do_plot_compl_bound}}{
        \draw [dashed, red, line width=0.2mm] (5.789e1, 1e-5) --  (5.789e1, 1) node [very near end, above, rotate=90, xshift=+3.5mm] {\tiny \SI{200}{\mega \hertz}};
    }{}
    
    \coordinate (bottom_left) at (3.5e1, 1e-5);
    \coordinate (top_right) at (1e3, 1);

    \begin{pgfonlayer}{background}

    \ifthenelse{\boolean{do_plot_compl_bound}}{
        \path[fill=customRed!3!white] (5.789e1, 1e-5) rectangle (1e3, 1);
    }{}

    \end{pgfonlayer}

    \coordinate (SelectedModel) at (axis cs:56.25, 0.000162292);
    
    \end{axis}

    \ifthenelse{\boolean{do_plot_sel_model}}{
        \draw[line width=0.2mm, black] (SelectedModel) circle [x radius=0.25cm, y radius=0.25cm];
    }{}

\end{tikzpicture}
	\caption{Results of design space exploration of the different equalization approaches. The maximal $\mathrm{MAC}_\mathrm{sym}$ to achieve the throughput of \SI{40}{\giga Bd} with a clock frequency of \SI{200}{\mega\hertz} is given by the vertical red line. The Pareto optimal models of the \ac{cnn}-based equalizer, the Volterra kernel, and the \ac{fir} filter are connected by the dotted, solid, and dashed lines respectively.}
	\label{fig:dse}
\end{figure}

}

The Pareto optimal models of each approach correspond to the most promising candidates for implementation since they provide the best trade-off between complexity and communication performance.

Further, the framework approximates the maximal $\mathrm{MAC}_\mathrm{sym}$ to achieve the required throughput $T_\mathrm{req}$ of \SI{40}{\giga Bd} based on the clock frequency $f_\mathrm{clk}$, and the available \acp{dsp} of our target device as follows: 
\begin{equation*}
   \mathrm{MAC}_\mathrm{sym, max} = \frac{\mathrm{DSP}_\mathrm{avail}}{T_\mathrm{req}} \cdot f_\mathrm{clk} \cdot 1.2 \; .
\end{equation*}

A factor of \num{1.2} is introduced since some arithmetic operations are implemented using \ac{lut} resources instead of \acp{dsp} which increases the total number of feasible \ac{mac} operations. Note that the equation is an approximation, mostly based on previous investigations and experiments, specifically designed for our hardware architecture. The equation might not be universally applicable to other hardware architectures or use cases. 

Previous experiments showed that a clock frequency above \SI{200}{\mega \hertz} often results in timing violations. Thus, in our design-space exploration, we set the limit for $\mathrm{MAC}_\mathrm{sym}$ to the value which corresponds to an approximated throughput of \SI{40}{\giga Bd} with a clock frequency of \SI{200}{\mega \hertz}.
\new{
\begin{figure}[b]
    \centering
	\tikzsetnextfilename{CNN_final_topology}
\begin{tikzpicture}[node distance=0.2,>=latex]
    \tikzset{near start abs/.style={xshift=.01cm}}
    \def\minWid{4.6cm}; \def\minHei{0.5cm}; \def\arrowLen{0.6cm};
    \def\boxFontSize{\footnotesize}


    \node (G_in) {};
    \node[block, rounded corners, draw=RPTU_DarkGreen, minimum width=\minWid, minimum height=\minHei, below=\arrowLen of G_in] (G_conv_1) {\footnotesize Conv1D: $K = 9$, $S = 8$, $P = 10$};
    \node[block, rounded corners, draw=RPTU_Violett, minimum width=\minWid, minimum height=\minHei, below=0.04cm of G_conv_1] (G_batchnorm_1) {\footnotesize Batchnorm};
    \node[block, rounded corners, draw=RPTU_LightGreen, minimum width=\minWid, minimum height=\minHei, below=0.04cm of G_batchnorm_1] (G_elu_1) {\footnotesize ReLU};
    
    \node[block, rounded corners, draw=RPTU_DarkGreen, minimum width=\minWid, minimum height=\minHei, below=\arrowLen of G_elu_1] (G_conv_2) {\footnotesize Conv1D: $K = 9$, $S = 1$, $P = 10$};
    \node[block, rounded corners, draw=RPTU_Violett, minimum width=\minWid, minimum height=\minHei, below=0.04cm of G_conv_2] (G_batchnorm_2) {\footnotesize Batchnorm};
    \node[block, rounded corners, draw=RPTU_LightGreen, minimum width=\minWid, minimum height=\minHei, below=0.04cm of G_batchnorm_2] (G_elu_2) {\footnotesize ReLU};

    \node[block, rounded corners, draw=RPTU_DarkGreen, minimum width=\minWid, minimum height=\minHei, below=\arrowLen of G_elu_2] (G_conv_3) {\footnotesize Conv1D: $K = 9$, $S = 2$, $P = 10$};
    \node[below=\arrowLen of G_conv_3] (G_out) {};

    \draw[-{Latex[length=2mm]}, thick] (G_in) -- node[midway, right, xshift=0.1cm, yshift=0.05cm] {} (G_conv_1.north);
    \draw[-{Latex[length=2mm]}, thick] (G_elu_1) -- node[midway, right, xshift=0.1cm, yshift=0.05cm] {\footnotesize $C = 5$} (G_conv_2);
    \draw[-{Latex[length=2mm]}, thick] (G_elu_2) -- node[midway, right, xshift=0.1cm, yshift=0.05cm] {\footnotesize $C = 5$} (G_conv_3);
    \draw[-{Latex[length=2mm]}, thick] (G_conv_3) -- node[midway, right, xshift=0.1cm, yshift=0.05cm] {\footnotesize $V_\mathrm{p} = 8$} (G_out);

\end{tikzpicture}
	\caption{Final topology of the \ac{cnn}-based equalizer with three layers, where $K$ corresponds to the kernel size, $S$ to the stride, $P$ to the padding, and $V_p$ to the symbols calculated in parallel}
	\label{fig:network_topology}
\end{figure}
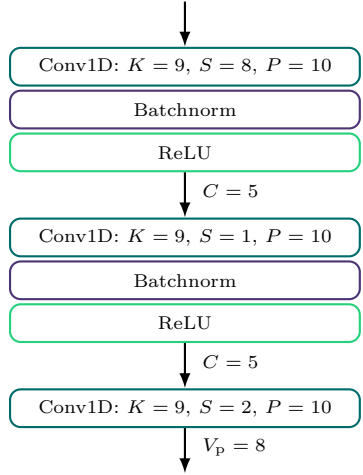
As a result, in Fig. \ref{fig:dse} we can see that for a fixed complexity in terms of \ac{mac} operations per symbol, the Pareto optimal \ac{cnn} models are outperforming the linear equalizer starting from a \ac{ber} of around $10^-2$ with respect to communication performance. Only when constrained to really low complexities of around \num{20} \ac{mac} operations per symbol, the linear equalizer provides a lower \ac{ber}. It can also be seen that the linear equalizer's performance saturates at a \ac{ber} of around \num{e-4}. This is probably the result of non-linear distortions of the optical fiber and \ac{cd} since those effects can not be fully compensated by a linear equalizer. In contrast to the \ac{fir} filter, the Volterra kernel introduces non-linearity in the equalization process. This way, similar to the \ac{cnn}, it is able to compensate for non-linear distortions. Thus, with sufficient complexity, the Volterra kernel provides a lower \ac{ber} than the \ac{fir} filter. However, for similar complexity, the Volterra kernel is outperformed by multiple \ac{cnn} configurations in terms of communication performance.}


For our hardware implementation, we select the configuration with the lowest \ac{ber} while satisfying our throughput requirements of \SI{40}{\giga Bd} with a clock frequency of \SI{200}{\mega \hertz}. This configuration is highlighted by the black circle and corresponds to a model with $V_p=8$, $L=3$, $K=9$, and $C=5$, as visualized in Fig. \ref{fig:network_topology}. In Fig. \ref{fig:dse}, we can see that the \ac{ber} achieved by a linear equalizer with the same complexity as the \ac{cnn} is around four times higher, while the Volterra's \ac{ber} is more than one order of magnitude higher. This shows that our \ac{cnn} topology is well suited for the equalization of the optical fiber channel.

\new{
\subsection{Performance for the Magnetic Recording Channel}

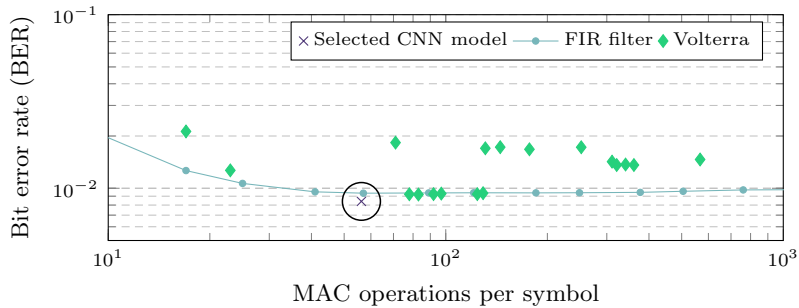
\begin{figure}[b]
	\centering
	\tikzsetnextfilename{proakis_cnn_vs_fir}

\begin{tikzpicture}
    [
    square/.style={regular polygon,regular polygon sides=4},
	/pgfplots/every mark/.append style={solid, mark size=3.5pt},
	/pgfplots/every axis/.append style={font=\small}, 
	/pgfplots/every tick label/.append style={font=\footnotesize},
	/pgfplots/every axis legend/.append style={nodes={right},font={\footnotesize},row sep=-2pt},
	]

	\begin{axis}[
            xmin=10,
            xmax=1000,
            ymin=0.005,
            ymax=0.1,
            xmode=log,
            ymode=log, 
            legend pos=north east,
            legend style={font=\footnotesize},
            legend columns=3,
            xlabel={MAC operations per symbol},
            ylabel={Bit error rate (BER)},
            width=0.8*\columnwidth,
            height=\columnwidth*0.35,
	        ymajorgrids,
            yminorgrids,
            grid style={line width=.01pt, densely dashed}
            ]

    \addplot[only marks, mark=x, color=RPTU_Violett, mark options={scale=1.2}] coordinates {(56.25, 0.0084)};
    \addlegendentry{Selected CNN model};

    \addplot [solid, mark=*, color=RPTU_GreenGray, mark options={scale=0.6}] table [x=MAC,y=BER]{data/DSE/fir_filter_proakis.txt}; 
    \addlegendentry{FIR filter};


    \addplot [only marks, mark=diamond*, color=RPTU_LightGreen, mark options={scale=1.2, fill=RPTU_LightGreen}] table [x=MAC,y=BER]{data/DSE/volterra_proakis.txt}; 
    \addlegendentry{Volterra};

    \coordinate (SelectedModel) at (axis cs:56.25, 0.0084);

    \draw[line width=0.2mm, black] (SelectedModel) circle [x radius=0.25cm, y radius=0.25cm];


    \end{axis}

\end{tikzpicture}
	\caption{Complexity and communication performance of the selected model as compared to conventional \ac{fir} filters and Volterra kernels for the magnetic recording channel}
	\label{fig:proakis_cnn_vs_fir}
\end{figure}


To show the applicability of the \ac{cnn}-based equalizer to different application scenarios, we train the \ac{cnn} selected in the design space exploration for the magnetic recording channel presented in Sec. \ref{sec:telephone_channel}. We model the \ac{snr} of the bad-quality channel with \SI{20}{dB}. 
Similar to the high-throughput channel, we compare \ac{fir}-filter-based equalizers and Volterra-based equalizers to our \ac{cnn} approach in terms of complexity and communication performance. 

The results are shown in Fig. \ref{fig:proakis_cnn_vs_fir}. The \ac{cnn} achieves a \ac{ber} of \num{8.4e-3} while the linear \ac{fir} filter's \ac{ber} with similar complexity is \num{9.6e-3}. Thus the gap between the two equalization approaches is much smaller as compared the optical fiber channel. This is reasonable since the main advantage of the \ac{cnn} lies in the compensation of non-linear distortions which are not present in the simulated magnetic recording channel. Thus, for the linear distortions of this channel, the \ac{fir} filter achieves comparable performance. However, the \ac{cnn} still slightly outperforms the conventional \ac{fir} filter with similar complexity in terms of communication performance.  The Volterra equalizer provides similar performance as the \ac{fir} filter, with the drawback of slightly higher complexity. The results show that even though the \ac{cnn} topology is selected based on the high-throughput channel, it still achieves sufficient performance for other channels. This further demonstrates the high flexibility of \ac{cnn}-based equalizers.  

}
\section{Quantization}
\label{sec:quant}

In addition to the network topology, another essential aspect of an efficient \ac{ann} implementation on resource-constrained devices is the quantization of weights and activations. In contrast to the \num{32}-bit floating-point format used in software, each value on the \ac{fpga} is represented in fixed-point format with arbitrary decimal and fractional width on the \ac{fpga}. 

To explore the quantization efficiently, we include an automated quantization approach in our framework, similar to the one proposed in \cite{nikolic2020}. Therefore, the loss function is modified to simultaneously learn the precision of each layer while optimizing the accuracy of the \ac{ann} during training. This is achieved by using a differentiable interpolation of the bit-widths, which allows to train them using backpropagation. Similar to \cite{nikolic2020}, we include a \ac{qlf} in the loss function, which determines how aggressively to quantize. This enables efficient exploration of the trade-off between bit width and communication performance. 
\new{

The quantization-aware loss can be described by the following equation: 
\begin{equation*}
    \mathrm{loss} = \frac{1}{S_{\mathrm{in}}} \sum_{i=1}^{S_{\mathrm{in}}} (y_i - x_i)^2 + \mathrm{QLF} \cdot \frac{\mathrm{B}_\mathrm{p} + \mathrm{B}_\mathrm{a}}{2}   \; ,
\end{equation*}
where $S_\mathrm{in}$ is the input sequence length, $\mathbf{y}=(y_1, y_2, \ldots)$ are the predicted symbols, $\mathbf{x}=(x_1, x_2, \ldots)$ are the transmitted symbols, $\mathrm{B}_\mathrm{p}$ is the average number of bits of the trainable parameters and $\mathrm{B}_\mathrm{a}$ is the average number of bits of the activations. 

In contrast to \cite{nikolic2020}, where an integer representation together with a scaling factor that is coupled to the bit-width of the values is learned, we adjust the algorithm to separately learn the integer and fraction width. This way, there is no need to scale the values in hardware during computation, as the numbers are directly represented by their corresponding fixed-point value and can directly be mapped to our hardware architecture. With the help of our framework, we perform a quantization analysis for the most promising \ac{cnn} model found in our design space exploration. 

Our training process for quantization can be divided into three steps:

\begin{enumerate}
    \item Full precision training: Perform training in full precision to find a well-initialized model for quantization,
    \item Bit-width-aware training: Train bit width of weights and activations and optimize communication performance simultaneously,
    \item Fine-tuning: Fix bit width and perform further training iterations to improve communication performance.
\end{enumerate}

\begin{figure}[b]
	\centerline{\tikzsetnextfilename{ActBits_vs_Iters}

\begin{tikzpicture}
    \begin{axis}[
            xmin=0, 
            xmax=15000, 
            ymin=0, 
            ymax=40, 
            xtick={0, 2000, 4000, 6000, 8000, 10000, 12000, 14000},
            xticklabel style={
                /pgf/number format/fixed,
                /pgf/number format/precision=0,
                /pgf/number format/fixed zerofill
            },
            xtick pos=bottom,
            scaled x ticks=false,
            xlabel={Training iterations},
            ylabel={Average activation bits},
            legend pos=north east,
            legend style={font=\scriptsize, yshift=-25pt},
            width=\columnwidth*0.9,
            height=\columnwidth*0.45
            ]

    \addplot [solid, thick, mark=none, color=RPTU_GreenGray] table [x=Iters,y=QLF_0_5]{data/Quantization/ActBits_vs_Iters.txt}; 
    \addlegendentry{QLF: 0.5};

    \addplot [solid, thick, mark=none, color=RPTU_LightBlue] table [x=Iters,y=QLF_00_5]{data/Quantization/ActBits_vs_Iters.txt}; 
    \addlegendentry{QLF: 0.05};

    \addplot [solid, thick, mark=none, color=RPTU_Violett] table [x=Iters,y=QLF_000_5]{data/Quantization/ActBits_vs_Iters.txt}; 
    \addlegendentry{QLF: 0.005};

    \addplot [solid, thick, mark=none, color=RPTU_LightGreen] table [x=Iters,y=QLF_0000_5]{data/Quantization/ActBits_vs_Iters.txt}; 
    \addlegendentry{QLF: 0.0005};

    \draw [dashed, RPTU_Red, thin] (2000,0) -- (2000,40);
    \draw [dashed, RPTU_Red, thin] (10000,0) -- (10000,40);

    \draw (1000, 36) circle (0.25cm);
    \node at (1000, 36) {1};
    
    \draw (6000, 36) circle (0.25cm);
    \node at (6000, 36) {2};

    \draw (12500, 36) circle (0.25cm);
    \node at (12500, 36) {3};

    \end{axis}
\end{tikzpicture}
}
	\caption{Course of the average activation bit width during the three phases of quantized training for different \acp{qlf}.}
	\label{fig:bits_vs_iters}
	\centerline{\tikzsetnextfilename{BER_vs_Iters}

\begin{tikzpicture}
    \begin{axis}[
            xmin=0, 
            xmax=15000, 
            ymax=2, 
            xtick pos=bottom,
            xtick={0, 2000, 4000, 6000, 8000, 10000, 12000, 14000},
            xticklabel style={
                /pgf/number format/fixed,
                /pgf/number format/precision=0,
                /pgf/number format/fixed zerofill
            },
            scaled x ticks=false,
            xlabel={Training iterations},
            ylabel={Bit error rate (BER)},
            ymode=log, 
            legend pos=north east,
            legend style={font=\scriptsize, yshift=-22pt},
            width=\columnwidth*0.9,
            height=\columnwidth*0.45
            ]

    \addplot [solid, thick, mark=none, color=RPTU_GreenGray] table [x=Iters,y=QLF_0_5]{data/Quantization/BER_vs_Iters.txt}; 
    \addlegendentry{QLF: 0.5};

    \addplot [solid, thick, mark=none, color=RPTU_LightBlue] table [x=Iters,y=QLF_00_5]{data/Quantization/BER_vs_Iters.txt}; 
    \addlegendentry{QLF: 0.05};

    \addplot [solid, thick, mark=none, color=RPTU_Violett] table [x=Iters,y=QLF_000_5]{data/Quantization/BER_vs_Iters.txt}; 
    \addlegendentry{QLF: 0.005};

    \addplot [solid, thick, mark=none, color=RPTU_LightGreen] table [x=Iters,y=QLF_0000_5]{data/Quantization/BER_vs_Iters.txt}; 
    \addlegendentry{QLF: 0.0005};

            
    \draw [dashed, RPTU_Red, thin] (2000,1e-5) -- (2000,2);
    \draw [dashed, RPTU_Red, thin] (10000,1e-5) -- (10000,2);

    \draw [dash dot, gray, thick] (0, 0.00014) -- (15000, 0.00014);

    \draw (1000, 0.8) circle (0.25cm);
    \node at (1000, 0.8) {1};
    
    \draw (6000, 0.8) circle (0.25cm);
    \node at (6000, 0.8) {2};

    \draw (12500, 0.8) circle (0.25cm);
    \node at (12500, 0.8) {3};
    
    \end{axis}
\end{tikzpicture}
}
	\caption{Course of the \ac{ber} during the three phases of quantized training for different \acp{qlf}. The \ac{ber} of the full precision model is given by the gray horizontal line.}
	\label{fig:ber_vs_iters}
\end{figure}

In Fig. \ref{fig:bits_vs_iters}, the training of the bit widths is shown for different \acp{qlf}.
In the first phase, the bit width is fixed to 32 bit, where 16 bits are used for the integer and 16 bits for the fractional part. In the second phase, the bit widths are reduced linearly until they saturate at different values for each \ac{qlf}. In the fine-tuning phase, there is a small increase in bit width again, since they get fixed to the next highest integer. 
The corresponding \ac{ber} of the \ac{cnn} is shown in Fig. \ref{fig:ber_vs_iters}. It can be seen that the \ac{ber} is reduced continuously for all \acp{qlf} until training iteration \num{4000}. Starting from this iteration, the low bit width starts to sacrifice the communication performance of the \acp{cnn}. Especially for a \ac{qlf} of \num{0.5} and \num{0.005}, the \ac{ber} increases dramatically. For a \ac{qlf} of \num{0.005} and \num{0.0005} there is only a slight increase in \ac{ber}. However, this increase can be compensated in the fine-tuning phase, where the quantized model nearly achieves the same \ac{ber} as the full precision model. As a result of the quantization-aware training, our \ac{fpga} hardware architecture is based on a model with around 13 bits for weights and 10 bits for activations with approximately the same communication performance as the full precision model.

}

\section{Hardware Architecture}
\label{sec:hardware_architecture}

In this section, the \ac{fpga} hardware architecture of our \ac{cnn}-based equalizer is presented. We mainly focus on the high-throughput requirements of the optical fiber channel, but also show how the architecture can be applied to other application scenarios. 

\subsection{High-Throughput Architecture}
\label{sec:high_throughput_architecture}
\begin{figure}[t]
	\centering
        \includegraphics[width=\columnwidth]{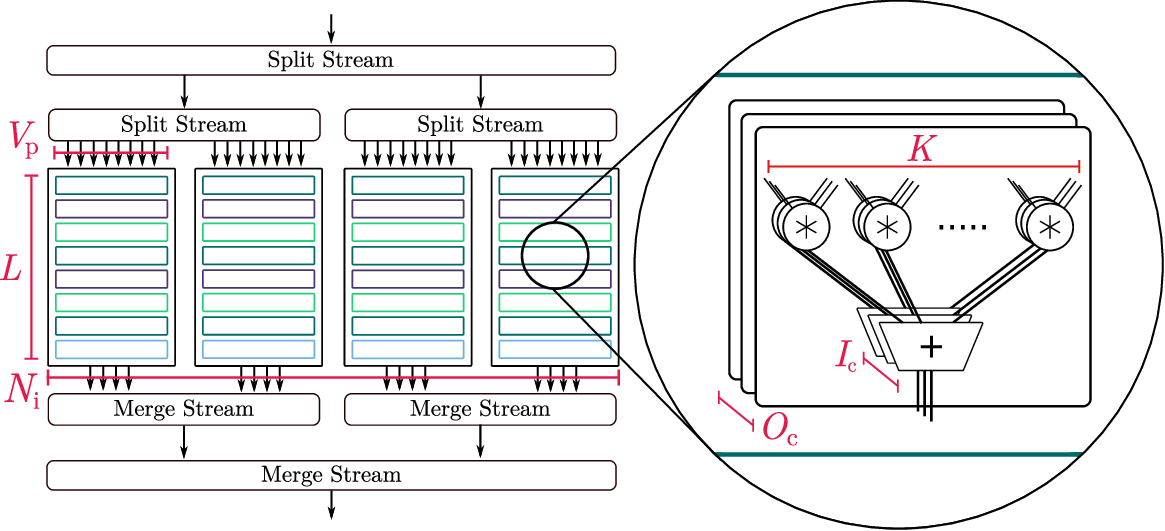}
	\caption{Applied levels of parallelism of our hardware architecture illustrated for four instances.}
        \vspace*{-10mm}
	\label{fig:parallel}
\end{figure}

The main target of our hardware implementation is to increase the throughput to meet the requirements of the \SI{40}{\giga Bd} optical communication channel. 
To satisfy the strict throughput requirements, it is essential to use the \ac{fpga}'s resources efficiently by increasing the utilization. Thus, the aim of our hardware architecture is to boost parallelism on all implementation levels. 
All those different levels of parallelism are illustrated in Fig. \ref{fig:parallel}. The first level of parallelism is based on our streaming hardware architecture, where each of the $L$ layers is implemented as an individual hardware instance. This way, the data is processed in a pipelined fashion, where each layer corresponds to a separate pipeline stage. Thus each layer can start its operation as soon as the first inputs are received which increases the throughput and the utilization of the available resources.  

The core of our hardware architecture is a custom convolutional layer that is also optimized for high parallelism. The convolution operation can be described by: 
\begin{equation}
\label{eq:conv}
    y_{o,j} = \sum_{i=0}^{I_c} \sum_{k=-\frac{K-1}{2}}^{\frac{K-1}{2}} x_{i,j+k} \cdot w_{i,o,k} \quad \mathlarger{\forall} o \in O_c    \; ,
\end{equation}
where $x$ is the input, $y$ the output, $w$ the kernel, $K$ the kernel size, $I_c$ the number of input channels and $O_c$ the number of output channels. From (\ref{eq:conv}), it can be seen that the convolutional layer offers multiple possibilities to apply spatial parallelism: on the level of input channels $I_c$, on the level of output channels $O_c$ and on the kernel level $K$. Our hardware architecture of the convolutional layer exploits all of those parallelization options to achieve maximal throughput, as shown on the right of Fig. \ref{fig:parallel}. Thus, it can achieve a throughput of one symbol per clock cycle. Since all of our layers are pipelined, this is also the throughput achieved by one hardware instance of the \ac{cnn}. 
Another level of parallelism that is exploited by our implementation is the number of \ac{cnn} instances $N_\mathrm{i}$. We place and connect multiple instances of the \ac{cnn} in one design to further boost the throughput. Therefore, the input is split into multiple streams, so each instance operates on a subset of the input sequence and produces a subset of the output sequence. Splitting and merging the sequence introduces further challenges, as explained later in Sec. \ref{sec:slo}. 
 In Fig. \ref{fig:parallel}, we can see that increasing parallelism is a major objective across multiple design layers: starting from the topology, where $V_\mathrm{p}$ symbols are calculated in parallel, over the pipelined layers $L$ and the parallelism with respect to $K$, $O_c$ and $I_c$ in the convolutional layer, up to the number of hardware instances $N_i$. This way several symbols are processed in parallel which is essential to reach the required throughput. In particular, the maximal throughput $T_\mathrm{max}$ in \si{\giga bit \per \second}  of our implementation is given as 
\begin{equation*}
    T_\mathrm{max} = N_\mathrm{i} \cdot V_\mathrm{p} \cdot f_\mathrm{clk}  \; .
\end{equation*}
We refer to maximal throughput here, as this throughput is only the theoretical upper limit, as explained later in Sec. \ref{sec:slo}. 
\new{
\subsection{Flexibility of Hardware Architecture}
As described in Sec. \ref{sec:high_throughput_architecture}, our implementation is well-suited for high-throughput scenarios. Additionally, due to the flexible design of our hardware architecture, it can also be applied to applications that operate at lower data rates. For those applications, reducing the power consumption or targeting low-cost \acp{fpga} might be more relevant than increasing throughput. To also satisfy those requirements, our hardware architecture allows for variable \acp{dop} based on the parallelization levels presented in Sec. \ref{sec:high_throughput_architecture}. 
One example of a channel with a limited maximal throughput is the Proakis-B magnetic recording channel as described in Sec. \ref{sec:telephone_channel}. 
In the following, we show how our hardware architecture is able to adapt to less strict throughput requirements. As an implementation platform, we select the low-cost \ac{fpga} Xilinx XC7S25-1CSGA324.

\begin{samepage}
For each hardware instance, parallelism can be applied to the input channels $\mathrm{DOP}_\mathrm{I}$, the output channels $\mathrm{DOP}_\mathrm{O}$, and to the kernel $\mathrm{DOP}_\mathrm{K}$. The final \ac{dop} is then given as: $\mathrm{DOP} = \mathrm{DOP}_\mathrm{I} \cdot \mathrm{DOP}_\mathrm{O} \cdot \mathrm{DOP}_\mathrm{K}$. However, the individual \acp{dop} are constrained by the hardware architecture as
\begin{align*}
    &\mathrm{I}_\mathrm{c} \equiv 0 \bmod \mathrm{DOP}_\mathrm{I}\mathrm{,}\\
    &\mathrm{O}_\mathrm{c} \equiv 0 \bmod \mathrm{DOP}_\mathrm{O}\mathrm{,}\\
    &\mathrm{DOP}_\mathrm{K} \in\{1, \mathrm{K}\}\mathrm{.}
\end{align*}
For our \ac{cnn} topology this results in $\mathrm{DOP} \in \{1, 5, 10, 25, 225\}$.
\end{samepage}
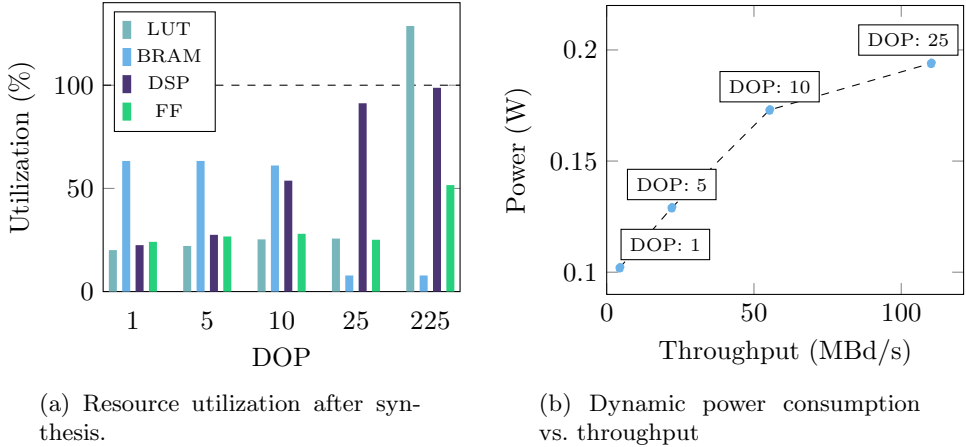
\begin{figure}[b]
    \begin{subfigure}[t]{0.48\columnwidth}
	\tikzsetnextfilename{Unrolling_Resources}

\begin{tikzpicture}
    \begin{axis}[
            xtick = data,
            xtick style={draw=none},
            ymin=0, 
            ymax=140, 
            xlabel={DOP},
            ylabel={Utilization (\%)},
            legend pos=north west,
            legend style={font=\footnotesize},
            width=\columnwidth,
            symbolic x coords={1, 5, 10, 25, 225},
            ybar,
            legend image code/.code={
            \draw [#1] (0cm,-0.1cm) rectangle (0.1cm,0.125cm); }
            ]


        \draw [dashed, black] (rel axis cs:0,0.7143) -- (rel axis cs:1,0.7143);
        
        \addplot [RPTU_GreenGray, fill=RPTU_GreenGray, bar width=3pt, draw=none] coordinates {(1,20.1) (5,22.1) (10,25.3) (25,25.7) (225,128.7)};
        
        \addplot [RPTU_LightBlue, fill=RPTU_LightBlue, bar width=3pt, draw=none] coordinates {(1,63.3) (5,63.3) (10,61.1) (25,7.8) (225,7.8)};

        \addplot [RPTU_Violett, fill=RPTU_Violett, bar width=3pt, draw=none] coordinates {(1,22.5) (5,27.5) (10,53.75) (25,91.25) (225,98.75)};

        \addplot [RPTU_LightGreen, fill=RPTU_LightGreen, bar width=3pt, draw=none] coordinates {(1,24.1) (5,26.7) (10,28.0) (25,25.1) (225,51.6)};

        \legend{LUT,BRAM,DSP,FF}
    \end{axis}
\end{tikzpicture}%
     \centering\captionsetup{width=.8\linewidth}%
	\caption{Resource utilization after synthesis.}
	\label{fig:unrolling_resources}
    \end{subfigure}\hspace{3mm}%
    \begin{subfigure}[t]{0.48\columnwidth}
	\tikzsetnextfilename{Power_vs_Throughput}

\begin{tikzpicture}
    \begin{axis}[
            xmin=0, 
            ymin=0.09, 
            ymax=0.22, 
            log ticks with fixed point,
            xlabel={Throughput (\si{\mega Bd \per \second})},
            ylabel={Power (\si{\watt})},
            width=\columnwidth,
            ]

        \addplot [dashed, mark=*, mark options={RPTU_LightBlue, scale=0.8}] table {
            4.429 0.102
            22.087 0.129
            55.317 0.173
            110.1625 0.194
        };

        \node [above, rectangle,draw, fill = white, xshift=6mm, yshift=1mm] at (axis cs:  4.429,  0.102) {\footnotesize DOP: \num{1}};
        \node [above, rectangle,draw, fill = white, yshift=1mm] at (axis cs:  22.087,  0.129) {\footnotesize DOP: \num{5}};
        \node [above, rectangle,draw, fill = white, yshift=1mm] at (axis cs:  55.317,  0.173) {\footnotesize DOP: \num{10}};
        \node [above, rectangle,draw, fill = white, xshift=-3mm, yshift=1mm] at (axis cs:  110.1625,  0.194) {\footnotesize DOP: \num{25}};

    \end{axis}
\end{tikzpicture}%
    \centering\captionsetup{width=.8\linewidth}%
	\caption{Dynamic power consumption vs. throughput}
	\label{fig:power_vs_tp}
    \end{subfigure}
    \caption{Resource utilization and power consumption vs. throughput on the Xilinx XC7S25-1CSGA324 for different \acp{dop}.}
\end{figure}
In Fig. \ref{fig:unrolling_resources}, the resource utilization on the target \ac{fpga} for different \acp{dop} is shown. It can be seen that primarily more \acp{lut} and \acp{dsp} are required with higher \ac{dop} since more \ac{mac} operations are performed in parallel. For a \ac{dop} of \num{255}, all available \acp{dsp} are used, thus the \ac{mac} operations are implemented using \acp{lut} which increases the \ac{lut} utilization above \SI{100}{\percent}. Further, Vivado HLS implements the trainable parameters using \acp{bram} for the smaller \acp{dop} and uses \ac{lut} resources as storage for the larger \acp{dop}. To summarize, Fig. \ref{fig:unrolling_resources} shows that our hardware architecture is also well suited for low-cost \acp{fpga} as it can be adapted to exploit the available hardware resources. 

In Fig. \ref{fig:power_vs_tp}, we show how the \ac{dop} influences the dynamic power consumption and the throughput of the implementation. A lower \ac{dop} results in fewer \ac{mac} operations per clock cycle leading to lower throughput and lower power consumption. It can be seen that one instance of the \ac{cnn} on the XC7S25-1CSGA324 can be adjusted to achieve a throughput in the range of \SI{4}{\mega \bit \per \second} to \SI{110}{\mega \bit \per \second} while the power ranges from \SI{0.1}{\watt} to \SI{0.2}{\watt}. 
}

\subsection{Stream Partitioning}
\label{sec:stream_partitioning}

\begin{figure}[tb]
	\centering
        \includegraphics[width=0.8\columnwidth]{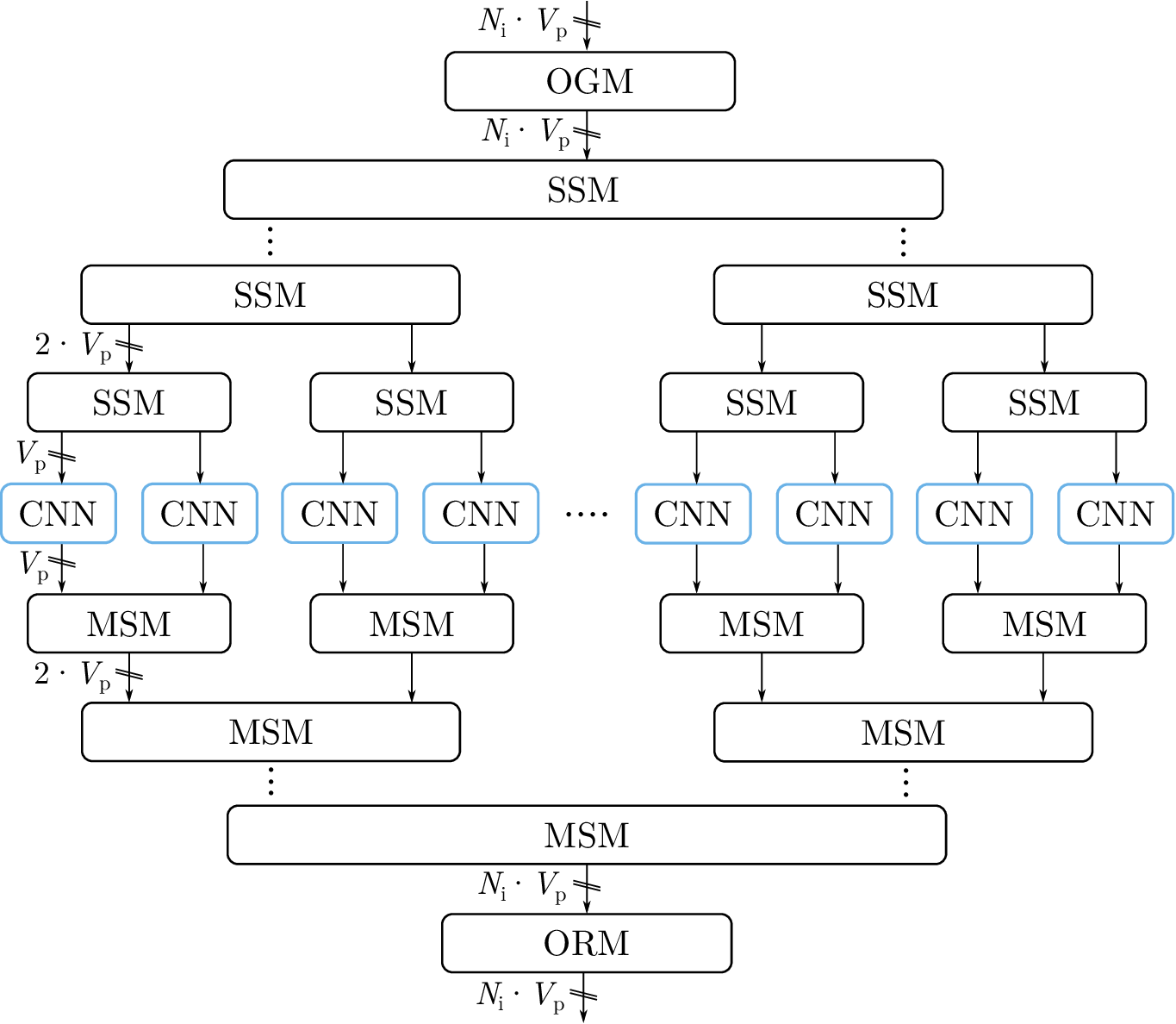}
	\caption{Partitioning of the input sequence across multiple \ac{cnn} instances.}
	\label{fig:split_merge}
\end{figure}

As explained in Sec. \ref{sec:high_throughput_architecture}, for the high-throughput scenario our architecture provides a high level of parallelism, especially with respect to the number of \ac{cnn} hardware instances. Splitting a stream of input symbols across those instances is not straightforward, since the \ac{isi} of the channel introduces an interdependence between consecutive symbols. Thus each \ac{cnn} instance needs to operate on a continuous sequence of input symbols. Therefore, we design a hardware module for splitting the input stream (\ac{ssm}) and a hardware module for merging the output streams (\ac{msm}), as shown in Fig. \ref{fig:split_merge}.

Each \ac{ssm} takes an input stream and splits it into two streams of equal length. Multiple of those modules are arranged hierarchically to feed data to every \ac{cnn} instance in a round-robin fashion. Due to the interdependence between consecutive symbols, splitting the input stream results in an increased \ac{ber} at the border region of each sequence. Thus, the \ac{ogm} adds an overlap to each sub-sequence. This way, the \ac{ber} is approximately constant for the complete stream.
Afterwards, the \acp{msm} combine the divided sequences into one output stream. Then, the overlap is discarded by the \ac{orm}. 
We arrange the \acp{ssm} and \acp{msm} in hierarchical fashion instead of just implementing one module which operates on $N_i$ streams. This improves the routability and timing of the design. Using only one module greatly increases the length of the paths from \ac{ssm} and \ac{msm} to each of the \ac{cnn} instances.  In combination with regions of high congestion, this results in an enormous net delay, limiting the achievable clock frequency. By introducing multiple \acp{ssm} and  \acp{msm}, the critical paths are shortened, which increases the achievable clock frequency and therefore the obtainable throughput.



\section{Sequence Length Optimization}
\label{sec:slo}

As described in Sec. \ref{sec:stream_partitioning}, each input sequence is divided into multiple sub-sequences of length $\ell_\mathrm{inst}$ which are forwarded to the individual instances. However, choosing an optimal $\ell_\mathrm{inst}$ under given application constraints is not straightforward.  On the one side, an overlap of symbols is added at the beginning and the end of each sub-sequence. From this point of view, maximizing $\ell_\mathrm{inst}$ results in the highest throughput since the overall overlap is minimized. On the other side, increasing $\ell_\mathrm{inst}$ also increases the latency of the equalizer, which may violate low-latency constraints in applications like high-frequency trading or telemedicine. Thus it is important to choose $\ell_\mathrm{inst}$ carefully, as described in the following.

\subsection{Timing Model}
\label{sec:timinig_model}
To optimize $\ell_\mathrm{inst}$, we perform a detailed timing analysis of our hardware architecture. First, we analyze how many overlap symbols need to be added at the beginning and end of each sequence to compensate for the \ac{ber} increase.  For a \ac{cnn}, the receptive field corresponds to the input symbols taken into account to predict each output. Thus, at the beginning and end of each sequence, half of the receptive field needs to be added as overlap. Based on the formula presented in \cite{Araujo2019}, the number of overlap symbols for our network topology is calculated as
\begin{equation*}
    o_\mathrm{sym} = \frac{(K - 1) \cdot (1 + V_\mathrm{p} \cdot (L - 1))}{2}  \; .
\end{equation*}
However, this overlap needs to be added before the first \ac{ssm} by the \ac{ogm}, where the stream has a width of $N_\mathrm{i} \cdot V_p$ and has to be dividable by $N_\mathrm{os}$, which equals to \num{2} in our case. Thus, the actual overlap can be calculated as
\begin{equation*}
    o_\mathrm{act} = \mathrm{nextEven}\Bigl(\left \lceil{\frac{o_\mathrm{sym}}{V_\mathrm{p} \cdot N_\mathrm{i}}}\right \rceil \Bigr) \cdot V_\mathrm{p} \cdot N_\mathrm{i}   \; .
\end{equation*}

Therefore, the actual sequence length that needs to be processed including overlap is given as
\begin{equation*}
    \ell_\mathrm{ol} = \ell_{\mathrm{inst}} + 2 \cdot o_\mathrm{act}    \; .
\end{equation*}

As a second step, we analyze how $\ell_\mathrm{ol}$ influences the time to fill the pipeline $t_\mathrm{init}$, afterwards, we explain how this affects the latency of each symbol.

\begin{figure}[t]
	\centering
	\vspace*{0.2cm}
        \includegraphics[width=0.9\columnwidth]{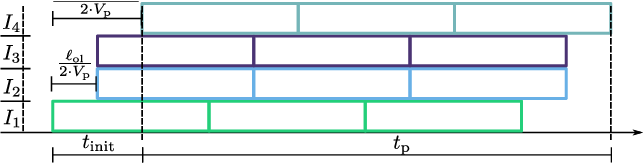}
	\caption{Illustration of the processing time $t_p$ and the time for filling the pipeline $t_\mathrm{init}$ for four \ac{cnn} instances}
	\label{fig:latency}
\end{figure}

In Fig. \ref{fig:latency}, we show how $\ell_\mathrm{ol}$ and therefore $\ell_\mathrm{inst}$ impact the total time $t$ to process one sequence. This time can be split into $t_\mathrm{init}$ and $t_\mathrm{p}$ and is illustrated for four instances. Since the width of the output streams of an \ac{ssm} is half the width of the input stream and the sequences of length $\ell_\mathrm{ol} / V_\mathrm{p}$ are written alternately to each output, the writing to the second output stream only starts after $\ell_\mathrm{ol} / (2 \cdot V_\mathrm{p})$ clock cycles. A similar behavior can be observed for each stage of the hierarchically arranged \acp{ssm}. Therefore, $t_{\mathrm{init}}$, corresponding to the time where the last \ac{cnn} instance starts processing, is given by
\begin{equation*}
    t_{\mathrm{init}} = \log_2(N_\mathrm{i}) \cdot \frac{\ell_\mathrm{ol}}{2 \cdot V_\mathrm{p} \cdot f_\mathrm{clk}}   \; .
\end{equation*}
Thus, it can be seen that $t_{\mathrm{init}}$ increases linearly with $\ell_{\mathrm{ol}}$ and $\ell_{\mathrm{inst}}$. Is it important to determine $t_\mathrm{init}$, as it directly influences the latency to process one symbol $\lambda_{\mathrm{sym}}$, which is given as the sum of the latency for splitting $\lambda_{\mathrm{spl}}$, processing $\lambda_{\mathrm{pro}}$ and merging $\lambda_{\mathrm{mer}}$. As the \ac{cnn} is fully parallelized and there is no stalling in the merging of streams, $\lambda_{\mathrm{pro}}$ and $\lambda_{\mathrm{mer}}$ are neglectable. In contrast, for splitting, a stream of higher width is converted into streams of lower width, which results in stalling and increased latency. The maximum symbol latency can therefore be approximated as the time $t_\mathrm{init}$ to fill the pipeline:
\begin{equation}
\label{eq:latency}
    \lambda_{\mathrm{sym}} \approx t_{\mathrm{init}} =  \frac{\log_2(N_\mathrm{i}) \cdot \ell_\mathrm{ol}}{2 \cdot V_\mathrm{p} \cdot f_\mathrm{clk}} = \frac{\log_2(N_\mathrm{i}) \cdot (\ell_\mathrm{inst} + 2 \cdot o_\mathrm{act})}{2 \cdot V_\mathrm{p} \cdot f_\mathrm{clk}}\; .
\end{equation}

From (\ref{eq:latency}), it can be seen that higher $\ell_{\mathrm{inst}}$ negatively impacts the symbol latency $\lambda_{\mathrm{sym}}$. From this point of view, it would be beneficial to set $\ell_{\mathrm{inst}}$ as small as possible. 

However, since $o_\mathrm{act}$ is fixed and is added to each sub-sequence of length $\ell_\mathrm{inst}$, the total number of symbols to process by the \ac{cnn} instances grows with shorter $\ell_\mathrm{inst}$. This is directly reflected in the processing time for one sequence of length $\ell_\mathrm{in}$, which is calculated as
\begin{equation*}
    t_p = \frac{\ell_\mathrm{in}}{\ell_\mathrm{inst} \cdot N_\mathrm{i}} \cdot \frac{\ell_\mathrm{inst} + 2 \cdot o_\mathrm{act}}{V_\mathrm{p} \cdot f_\mathrm{clk}} = \frac{\ell_\mathrm{in}}{N_\mathrm{i} \cdot V_\mathrm{p} \cdot f_\mathrm{clk}} \cdot \Bigl(1 + \frac{2 \cdot o_\mathrm{act}}{\ell_\mathrm{inst}}\Bigr)   \; .
\end{equation*}

The processing time is inversely proportional to the net throughput:
\begin{equation}
\label{eq:net_throughput}
T_\mathrm{net} = \frac{\ell_\mathrm{in}}{t_p} = \frac{N_\mathrm{i} \cdot V_\mathrm{p} \cdot f_\mathrm{clk}}{1 + \frac{2 \cdot o_\mathrm{act}}{\ell_\mathrm{inst}}}   \; .
\end{equation}

Thus, the net throughput grows with larger $\ell_\mathrm{inst}$.
In summary, both the symbol latency $\lambda_{\mathrm{sym}}$ and the throughput $T_\mathrm{net}$ increase with $\ell_\mathrm{inst}$, therefore a trade-off exists when optimizing for throughput and latency. 

\subsection{Optimization Framework}
\label{sec:optimization_framework}

\begin{figure}[!b]
	\centering
        \includegraphics[width=0.6\columnwidth]{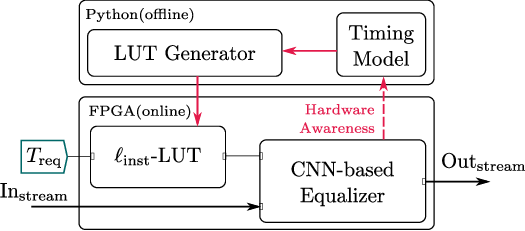}
	\caption{Illustration of framework used to optimize sequence length per instance $\ell_\mathrm{inst}$.}
	\label{fig:framework}
\end{figure}

Based on the equations presented in the previous section, we propose a framework to select the best $\ell_\mathrm{inst}$ for the given application requirements. In our case, the throughput is a hard constraint that needs to be satisfied, while latency is an objective we want to minimize. Thus, the framework selects the minimal $\ell_\mathrm{inst}$ which satisfies the throughput requirements. 

The framework is shown in Fig. \ref{fig:framework}. Part of the framework, in particular the lookup table which maps the required throughput $T_\mathrm{req}$ to the optimal sub-sequence length $\ell_\mathrm{inst}$, is implemented as a module on the \ac{fpga}. This way, the best $\ell_\mathrm{inst}$ can be selected during runtime individually for each sequence to process. The lookup table is provided by a lookup-table-generator based on the timing model presented in Sec. \ref{sec:timinig_model}. The timing model is derived from a detailed analysis of the hardware architecture and therefore introduced hardware awareness to the LUT-Generator. This information is fed back to the hardware by selecting $\ell_\mathrm{inst}$. This way, our framework is also based on a cross-layer design methodology.

\section{Results}

In the following, the results of our timing model, our \ac{ht} \ac{fpga} implementation, and our \ac{lp} \ac{fpga} implementation are presented. 
The main goal of our high-throughput hardware implementation is to achieve the throughput required for equalizing the \SI{40}{\giga Bd} optical communication channel. As the channel data is captured with an upsampling factor of $N_\mathrm{os} = 2$ at the receiver, this corresponds to \SI{80}{\giga samples \per \second} at the input of our equalizer. 
To allow for such high data rate, we choose the least complex \ac{cnn} model still satisfying our \ac{ber} requirements in combination with multiple levels of parallelism of our \ac{fpga} architecture. Furthermore, we make use of our timing model and framework to estimate the number of instances needed to achieve the required throughput. 
As a hardware platform for the high throughput channel, we select the high-performance \ac{fpga} Xilinx XCVU13-P with a huge amount of available resources, to allow for extremely high parallelism.
In contrast, for the \ac{lp} \ac{fpga} implementation of the equalizer for the magnetic recording channel, the Xilinx XC7S25-1CSGA324 \ac{fpga} is used. 
For both implementations, Vitis HLS in combination with Vivado 2022.2 is used. 

\subsection{Timing Model Validation}

\begin{figure}[bt]
	\centerline{\tikzsetnextfilename{TimingModelLatency}

\begin{tikzpicture}
    \begin{axis}[
            xmin=0, 
            xmax=9000, 
            ymin=0, 
            samples=200,
            xtick={0, 2000, 4000, 6000, 8000},
            xticklabel style={
                /pgf/number format/fixed,
                /pgf/number format/precision=0,
                /pgf/number format/fixed zerofill
            },
            scaled x ticks=false,
            xlabel={$\ell_\mathrm{inst}$},
            ylabel={$\lambda_\mathrm{sym}$ (\si{\micro \second})},
            legend pos=north west,
            legend style={font=\scriptsize},
            width=0.5*\columnwidth,
            ]

      \addplot[domain=64:10000, black, thick, dotted] ({x}, {(4 * (x + 2 * 256)) / (2 * 8 * 0.2 * 1000)}); 

      \addlegendentry{\num{16} Instances};  

      \addplot[domain=64:10000, black, thick, dashed] ({x}, {(5 * (x + 2 * 512)) / (2 * 8 * 0.2 * 1000)}); 

      \addlegendentry{\num{32} Instances};  

      \addplot[domain=64:10000, black, thick, loosely dashdotted] ({x}, {(6 * (x + 2 * 1024)) / (2 * 8 * 0.2 * 1000)}); 

      \addlegendentry{\num{64} Instances};  
      
        \addplot [only marks, mark=x, mark options={RPTU_LightGreen, scale=1.6, thick}] table {
            1024 2.24
            2048 3.52
            4096 6.08
            8192 11.2
        };

        \addplot [only marks, mark=x, mark options={RPTU_LightGreen, scale=1.6, thick}] table {
            1024 3.425
            2048 5.005
            4096 8.205
            8192 14.605
        };

    \end{axis}
\end{tikzpicture}
\quad 

\tikzsetnextfilename{TimingModelThroughput}

\begin{tikzpicture}
    \begin{axis}[
            xmin=0, 
            xmax=9000, 
            ymin=0, 
            ymax=110,
            samples=200,
            xtick={0, 2000, 4000, 6000, 8000},
            xticklabel style={
                /pgf/number format/fixed,
                /pgf/number format/precision=0,
                /pgf/number format/fixed zerofill
            },
            scaled x ticks=false,
            extra y ticks={80},
            extra y tick labels={\textcolor{customRed}{80}},
            xlabel={$\ell_\mathrm{inst}$},
            ylabel={$T_\mathrm{net}$ (\si{\giga bit \per \second})},
            width=0.5*\columnwidth,
            ]

      \addplot[domain=64:10000, black, thick, dotted]  ({x}, {25.6 / (1 + 512 / x)}); 
      \draw [thick, loosely dashed, RPTU_LightBlue] (0,25.6) -- (10000,25.6);

      \addplot[domain=64:10000, black, thick, dashed]  ({x}, {51.2 / (1 + 1024 / x)}); 
      \draw [thick, loosely dashed, RPTU_LightBlue] (0,50.2) -- (10000,50.2);

      \addplot[domain=64:10000, black, thick, loosely dashdotted]  ({x}, {102.4 / (1 + 2048 / x)}); 
      \draw [thick, loosely dashed, RPTU_LightBlue] (0,100.4) -- (10000, 100.4);

      \draw [-, RPTU_Red] (0,80) -- (10000, 80);

        \addplot [only marks, mark=x, mark options={RPTU_LightGreen, scale=1.6, thick}] table {
            1024 17.07
            2048 20.48
            4096 22.76
            8192 24.09
        };

        \addplot [only marks, mark=x, mark options={RPTU_LightGreen, scale=1.6, thick}] table {
            1024 25.6
            2048 34.13
            4096 40.96
            8192 45.51
        };
      
    \end{axis}
\end{tikzpicture}
}
	\caption{Left: Plot of the sequence length $\ell_\mathrm{inst}$ against the symbol latency $\lambda_\mathrm{sym}$.\\ Right: Plot of the sequence length $\ell_\mathrm{inst}$ against the net throughput $T_\mathrm{net}$. On the right, the maximal theoretical throughput is given by the horizontal dashed lines}
	\label{fig:slt_vs_t}
\end{figure}
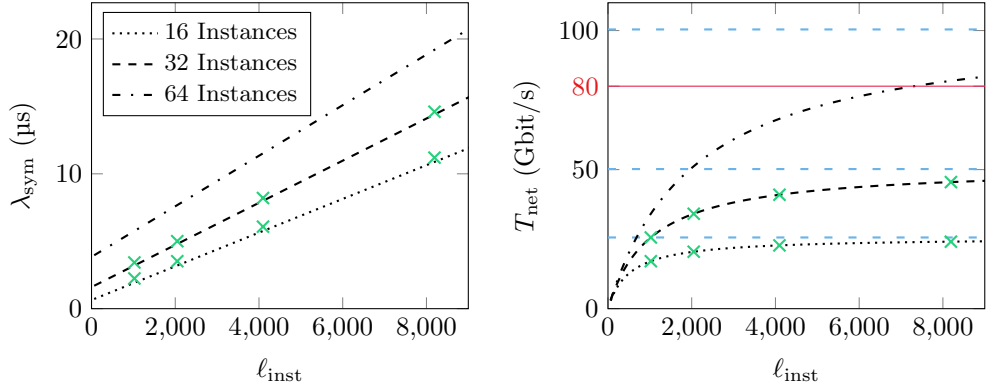

In the following, we validate the correctness of our timing model by comparing it to real timing measurements. Moreover, we evaluate how many \ac{cnn} instances are needed to achieve a throughput of \SI{80}{\giga samples \per \second} with a clock frequency of \SI{200}{\mega \hertz}. In Fig. \ref{fig:slt_vs_t} we show how $\ell_\mathrm{inst}$ influences the symbol latency $\lambda_\mathrm{sym}$ and the net throughput $T_\mathrm{net}$. 
The blue stars are based on simulations of the hardware, while the black graphs correspond to our timing model. The horizontal lines of the throughput plot give the maximal theoretical throughput $T_\mathrm{max}$ as $\ell_\mathrm{inst}\to\infty$.

It can be seen that both the latency as well as the throughput increase with a higher number of instances $N_\mathrm{i}$. While the latency grows linearly with the sequence length $\ell_{\mathrm{inst}}$, the throughput saturates for $\ell_\mathrm{inst}\to\infty$. The gap between $T_\mathrm{max}$ and $T_\mathrm{net}$ increases with $N_\mathrm{i}$ for a fixed $\ell_{\mathrm{inst}}$, which shows that it is necessary to select a larger $\ell_{\mathrm{inst}}$ when increasing $N_\mathrm{i}$ to reduce the influence of the overlap symbols. 
Further, it is shown that our equations are close to the actual measurements. In particular, the difference between measurements and model is only around \SI{6}{\percent} for latency and \SI{0.1}{\percent} for throughput, which validates the accuracy of our timing model. Thus, based on our model, we can reliably predict that at least \num{64} instances are needed to achieve the required throughput of \SI{80}{\giga samples \per \second}.
In addition, it can be seen that $\ell_\mathrm{inst}$ has a significant impact on $T_\mathrm{net}$. Thus,  $\ell_\mathrm{inst}$ is an important factor to consider for increasing performance, which justifies the use of our framework to satisfy throughput requirements.

\subsection{High-Throughput Implementation Results}
\addtolength{\tabcolsep}{3pt}    
\begin{table}[!b]
\centering
\caption{Utilization}
\label{tab:utilization}
\begin{tabular}{cc|cc|cc|cc}
\toprule
\multicolumn{2}{c}{LUT} & \multicolumn{2}{c}{FF} & \multicolumn{2}{c}{DSP} & \multicolumn{2}{c}{BRAM} \\

\midrule

\%         & absolute         & \%        & absolute        & \%         & absolute        & \%         & absolute         \\
\num{68.06}      &   \num{1176156}        &   \num{30.39}        &   \num{1050179}         &    \num{78.52}        &  \num{9648}          &   \num{78.79}         &   \num{2118}       \\  
           
\bottomrule         

\end{tabular}
\end{table}
\addtolength{\tabcolsep}{-3pt}

Based on our timing model, we know that at least \num{64} instances are needed to achieve the required throughput. As a result of our detailed design space exploration and our advanced hardware architecture, we are actually able to place \num{64} parallel instances of the \ac{cnn} model presented in Sec. \ref{sec:topology} on the board. The instances are connected by \num{63} \acp{ssm} and \acp{msm}, respectively, to compose a common input and output stream. The maximal achievable clock frequency for the design is \SI{200}{\mega \hertz}. Thus, the maximal throughput is given by:
\begin{equation*}
    T_\mathrm{max} = N_i \cdot V_p \cdot f_\mathrm{clk} = 64 \cdot \SI{8}{samples} \cdot \SI{200}{\mega \hertz} = \SI{102}{\giga samples \per \second} \mathrel{\widehat{=}} \SI{51}{\giga Bd} \; .
\end{equation*}

Based on our framework presented in Sec. \ref{sec:optimization_framework}, the minimal $\ell_\mathrm{inst}$ to achieve a net throughput of \SI{80}{\giga sym \per \second} is determined, which is \num{7320}. Choosing this sequence length results in the minimal symbol latency of only \SI{17.5}{\micro \second}, while satisfying the throughput requirements. 

In the following, we show the resource usage of our hardware architecture on the Xilinx XCVU13-P \ac{fpga}. In Table \ref{tab:utilization}, we give the utilization of \acp{lut}, \ac{ff}, \ac{dsp}, and \ac{bram} after place and route. It can be seen that the resources with the highest utilization are \acp{dsp} and \acp{bram}. \acp{dsp} are utilized for the \ac{mac} operation in the convolutional layers, while the \acp{bram} are mainly used for splitting and merging the input streams. Increasing the number of instances further to achieve higher utilization results in routing congestion and a lower clock frequency, eventually reducing the achievable throughput. 
\new{
\subsection{Platform Comparison}

In the following, we compare the throughput, latency, and power consumption achieved by our \ac{ht} and \ac{lp} \ac{fpga} implementations to other hardware platforms. As platforms we select the Nvidia RTX 2080 Ti high-performance \ac{gpu}, the Nvidia AGX Xavier embedded \ac{gpu}, and the Intel Core i9-9900KF high-performance \ac{cpu}. For the \ac{cpu} and the \ac{gpu}, we increase the batch size and therefore the \ac{spb} up to the point where the throughput does not improve further. For the \acp{fpga}, the \ac{spb} is fixed by the hardware architecture to \num{512} for the \ac{ht} implementation and to \num{8} for the \ac{lp} implementation. For fair comparison, for the \acp{gpu} we do not only evaluate the standard PyTorch implementation but also provide results for an optimized \ac{gpu} implementation. This implementation is based on the Nvidia library TensorRT which builds a highly optimized model for faster inference. The optimizations performed by TensorRT include quantization, layer and tensor fusion, and kernel tuning.

\subsubsection{Throughput}

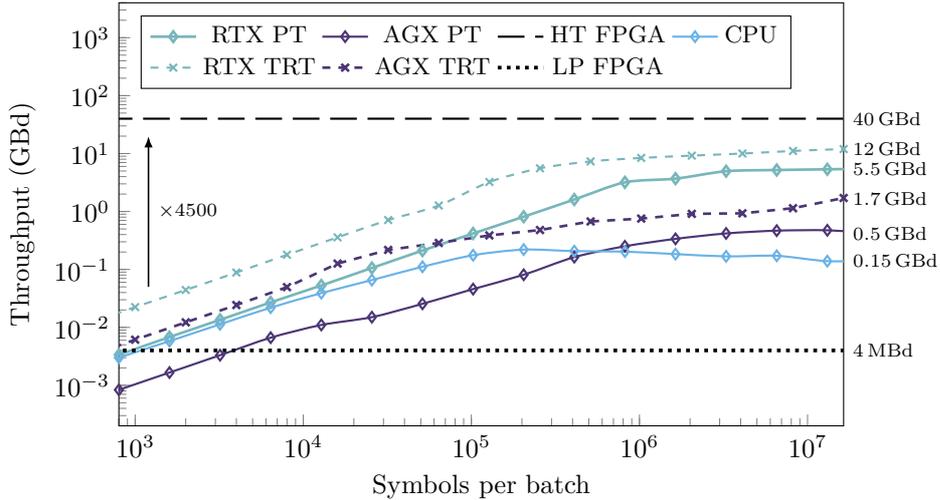
\begin{figure}[!t]
	\centerline{\tikzsetnextfilename{throughput_plot}

\begin{tikzpicture}
    \begin{axis}[
            xmode=log,
            ymode=log,
            log basis x={10},
            log basis y={10},
            xmin=800, 
            xmax=16384000, 
            ymin=2e-4, 
            tick pos=left,
            ymax=4000, 
            xlabel={Symbols per batch},
            ylabel={Throughput (\si{\giga Bd})},
            legend pos=north west,
            legend columns=4,
            legend style={font=\scriptsize},
            width=0.85*\columnwidth,
            height=\columnwidth*0.55] 

    \addplot [solid, smooth, mark=diamond,  mark options={solid}, color=RPTU_GreenGray, line width=1pt] table [x=symbols, y=throughput]{data/RTX_PyTorch_measurements.txt}; 
    \addlegendentry{RTX PT};

    \addplot [solid, smooth, mark=diamond,  mark options={solid}, color=RPTU_Violett, line width=0.8pt] table [x=symbols, y=throughput]{data/AGX_PyTorch_measurements.txt}; 
    \addlegendentry{AGX PT};

    \addplot[dash pattern={on 10pt off 4pt}, mark=none, black, line width=0.8pt, domain=400:16384000] {40.0};
    \addlegendentry{HT FPGA};

   \addplot [solid, smooth, mark=diamond,  mark options={solid}, color=RPTU_LightBlue, line width=0.8pt, domain=400:16384000] table [x=symbols, y=throughput]{data/i9_measurements.txt}; 
    \addlegendentry{CPU};

    \addplot [dashed, smooth, mark=x,  mark options={solid}, color=RPTU_GreenGray, , line width=0.8pt] table [x=symbols, y=throughput]{data/RTX_TRT_measurements.txt}; 
    \addlegendentry{RTX TRT};

    \addplot [dashed, mark=x,  mark options={solid}, color=RPTU_Violett, line width=1pt] table [x=symbols, y=throughput]{data/AGX_TRT_measurements.txt}; 
    \addlegendentry{AGX TRT};

    \addplot[dotted, mark=none, black, line width=1.4pt, domain=400:16384000] {0.004};
    \addlegendentry{LP FPGA};

     \draw [-{Latex[length=1.5mm]}, black, line width=0.6pt] (1200, 0.05)  --  node[right]{\footnotesize $\times 4500$} (1200, 40 - 20);

    \coordinate (htFPGA_max) at (axis cs:16384000, 40);
    \coordinate (rtxTRT_max) at (axis cs:16384000, 12);
    \coordinate (rtxPT_max) at (axis cs:16384000, 5.5);
    \coordinate (agxTRT_max) at (axis cs:16384000, 1.7112);
    \coordinate (agxPT_max) at (axis cs:16384000, 0.42);
    \coordinate (cpu_max) at (axis cs:16384000, 0.145);
    \coordinate (lpFPGA_max) at (axis cs:16384000, 0.004);

    \end{axis}

    \node[anchor=west] at (htFPGA_max) {\fontsize{7}{12}\selectfont \SI{40}{\giga Bd}};
    \node[anchor=west] at (rtxTRT_max) {\fontsize{7}{12}\selectfont \SI{12}{\giga Bd}};
    \node[anchor=west] at (rtxPT_max) {\fontsize{7}{12}\selectfont \SI{5.5}{\giga Bd}};
    \node[anchor=west] at (agxTRT_max) {\fontsize{7}{12}\selectfont \SI{1.7}{\giga Bd}};
    \node[anchor=west] at (agxPT_max) {\fontsize{7}{12}\selectfont \SI{0.5}{\giga Bd}};
    \node[anchor=west] at (cpu_max) {\fontsize{7}{12}\selectfont \SI{0.15}{\giga Bd}};
    \node[anchor=west] at (lpFPGA_max) {\fontsize{7}{12}\selectfont \SI{4}{\mega Bd}};

\end{tikzpicture}
}
	\caption{Comparison of the throughput of the \ac{cnn}-based equalizer, running on \ac{gpu}, \ac{cpu} and \ac{fpga}. PT refers to the PyTorch and TRT refers to the TensorRT models.}
	\label{fig:gpu_cpu_throughput}
\end{figure}

In Fig. \ref{fig:gpu_cpu_throughput}, the throughput of the different platforms is compared. For similar \ac{spb}, the \ac{ht} \ac{fpga} outperforms the \ac{cpu} and all \ac{gpu} implementations by orders of magnitude. For instance, the \ac{ht} \ac{fpga} achieves a \num{4500} times higher throughput than the RTX TensorRT model for 400 \ac{spb}. The throughput of the \ac{lp} \ac{fpga} is in the same order of magnitude as the AGX TensorRT model, the RTX PyTorch model, and the \ac{cpu} for $\mathrm{\ac{spb}} < 1000$. However, for large \ac{spb}, the throughput of the high-performance platforms is much higher than that of the \ac{lp} \ac{fpga}. 

\new{
It can also be seen that TensorRT highly optimizes the throughput of the \ac{cnn} models on \ac{gpu}. Especially for low batch sizes, the throughput is around one order of magnitude higher as compared to the PyTorch model for both the RTX TensorRT model and the AGX Tensor RT model. For all implementations, the throughput increases linearly for low \ac{spb} and saturates for high \ac{spb}. The \ac{fpga}'s throughput is constant over the whole range of \ac{spb} since the parallelization is fixed by the hardware architecture. Thus, parallelization does not increase with batch size but each batch is calculated sequentially on the \ac{fpga}. The highest throughput achieved by the conventional platforms is \SI{12}{\giga Bd} by the RTX TensorRT implementation. However, even for high \ac{spb}, the \ac{fpga} outperforms the RTX TensorRT model by $10$ times while the high-performance \ac{cpu} is outperformed by more than two orders of magnitude.}

\subsubsection{Latency}

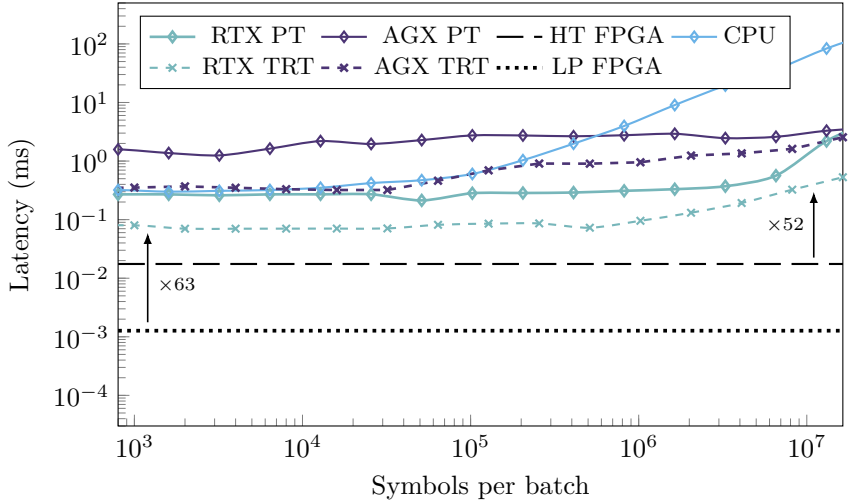
\begin{figure}[!t]
	\centerline{\tikzsetnextfilename{latency_plot}

\begin{tikzpicture}
       \begin{axis}[
            xmode=log,
            ymode=log,
            log basis x={10},
            log basis y={10},
            xmin=800, 
            xmax=16384000, 
            ymin=0.3e-4, 
            tick pos=left,
            ymax=500, 
            xlabel={Symbols per batch},
            ylabel={Latency (\si{\milli \second})},
            legend pos=north west,
            legend columns=4,
            legend style={font=\scriptsize},
            width=0.85*\columnwidth,
            height=\columnwidth*0.55
            ]

    \addplot [solid, smooth, mark=diamond,  mark options={solid}, color=RPTU_GreenGray, line width=1pt] table [x=symbols, y=latency]{data/RTX_PyTorch_measurements.txt}; 
    \addlegendentry{RTX PT};

    \addplot [solid, smooth, mark=diamond,  mark options={solid}, color=RPTU_Violett, line width=0.8pt] table [x=symbols, y=latency]{data/AGX_PyTorch_measurements.txt}; 
    \addlegendentry{AGX PT};

    \addplot[dash pattern={on 10pt off 4pt}, domain=800:104857600, mark=none, black, line width=0.8pt] {0.0175};
    \addlegendentry{HT FPGA};

   \addplot [solid, smooth, mark=diamond,  mark options={solid}, color=RPTU_LightBlue, line width=0.8pt] table [x=symbols, y=latency]{data/i9_measurements.txt}; 
    \addlegendentry{CPU};

    \addplot [dashed, smooth, mark=x,  mark options={solid}, color=RPTU_GreenGray, , line width=0.8pt] table [x=symbols, y=latency]{data/RTX_TRT_measurements.txt}; 
    \addlegendentry{RTX TRT};

    \addplot [dashed, mark=x,  mark options={solid}, color=RPTU_Violett, line width=1pt] table [x=symbols, y=latency]{data/AGX_TRT_measurements.txt}; 
    \addlegendentry{AGX TRT};

    \addplot[dotted, domain=800:104857600, mark=none, black, line width=1.4pt] {0.00127};
    \addlegendentry{LP FPGA};

     \draw [-{Latex[length=1.5mm]}, black, line width=0.6pt] (1200, 0.00127 + 0.5E-3)  --  node[right, yshift=-3pt]{\footnotesize $\times 63$} (1200, 	0.06);

     \draw [-{Latex[length=1.5mm]}, black, line width=0.6pt] (11000000, 0.0175 + 0.5E-2)  --  node[left]{\footnotesize $\times 52$} (11000000, 	0.3);

    \end{axis}
\end{tikzpicture}
}
	\caption{Comparison of the latency of the \ac{cnn}-based equalizer, running on \ac{gpu}, \ac{cpu} and \ac{fpga}. PT refers to the PyTorch and TRT refers to the TensorRT models.}
	\label{fig:gpu_cpu_latency}
\end{figure}

In Fig. \ref{fig:gpu_cpu_latency}, the latency of the different implementations is compared. It can be seen that even for low \ac{spb}, the latency of the \acp{gpu} and \ac{cpu} is more than one order of magnitude higher than that of the \ac{lp} \ac{fpga} and around $5\times$ higher than that of the \ac{ht} \acp{fpga}. This gap increases for higher \ac{spb} up to a factor of \num{52} between the \ac{ht} \ac{fpga} and the Nvidia AGX with the TensorRT model. As already seen in the throughput plot, TensorRT provides improved performance as compared to the default PyTorch model. Specifically, it decreased the latency of the models by around one order of magnitude.

\subsubsection{Power}

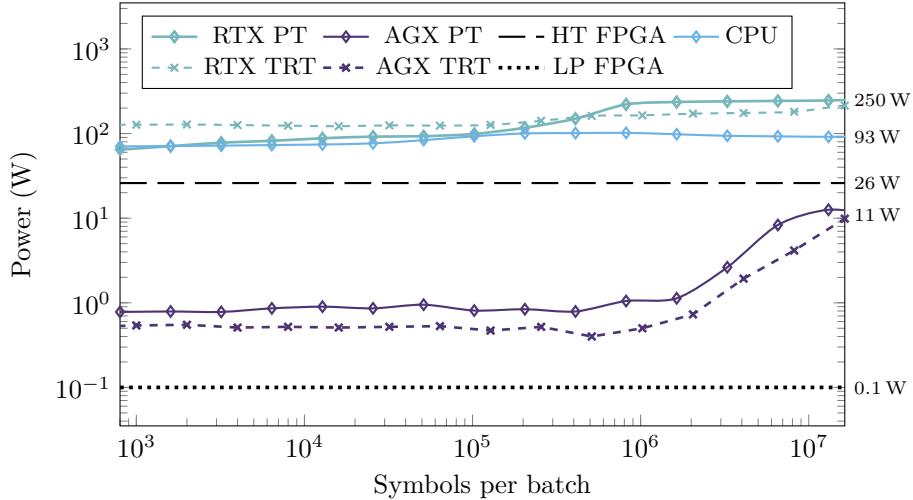
\begin{figure}[!t]
	\centerline{\tikzsetnextfilename{power_plot}

\begin{tikzpicture}
       \begin{axis}[
            xmode=log,
            ymode=log,
            log basis x={10},
            log basis y={10},
            xmin=800, 
            xmax=16384000, 
            ymax=3500, 
            xtick pos=bottom,
            xlabel={Symbols per batch},
            ylabel={Power (\si{\watt})},
            legend pos=north west,
            legend columns=4,
            legend style={font=\scriptsize},
            width=0.85*\columnwidth,
            height=\columnwidth*0.55
            ]


    \addplot [solid, smooth, mark=diamond,  mark options={solid}, color=RPTU_GreenGray, line width=1pt] table [x=symbols, y=power]{data/RTX_PyTorch_measurements.txt}; 
    \addlegendentry{RTX PT};

    \addplot [solid, smooth, mark=diamond,  mark options={solid}, color=RPTU_Violett, line width=0.8pt] table [x=symbols, y=power]{data/AGX_PyTorch_measurements.txt}; 
    \addlegendentry{AGX PT};

    \addplot[dash pattern={on 10pt off 4pt}, domain=800:104857600, mark=none, black, line width=0.8pt] {26};
    \addlegendentry{HT FPGA};

   \addplot [solid, smooth, mark=diamond,  mark options={solid}, color=RPTU_LightBlue, line width=0.8pt] table [x=symbols, y=power]{data/i9_measurements.txt}; 
    \addlegendentry{CPU};

    \addplot [dashed, smooth, mark=x,  mark options={solid}, color=RPTU_GreenGray, , line width=0.8pt] table [x=symbols, y=power]{data/RTX_TRT_measurements.txt}; 
    \addlegendentry{RTX TRT};

    \addplot [dashed, mark=x,  mark options={solid}, color=RPTU_Violett, line width=1pt] table [x=symbols, y=power]{data/AGX_TRT_measurements.txt}; 
    \addlegendentry{AGX TRT};

    \addplot[dotted, domain=800:104857600, mark=none, black, line width=1.4pt] {0.1};
    \addlegendentry{LP FPGA};

    \coordinate (htFPGA_max) at (axis cs:16384000, 26);
    \coordinate (rtxTRT_max) at (axis cs:16384000, 250);
    \coordinate (agxTRT_max) at (axis cs:16384000, 11);
    \coordinate (cpu_max) at (axis cs:16384000, 93);
    \coordinate (lpFPGA_max) at (axis cs:16384000, 0.1);

    \end{axis}

    \node[anchor=west] at (htFPGA_max) {\fontsize{7}{12}\selectfont \SI{26}{\watt}};
    \node[anchor=west] at (rtxTRT_max) {\fontsize{7}{12}\selectfont \SI{250}{\watt}};
    \node[anchor=west] at (agxTRT_max) {\fontsize{7}{12}\selectfont \SI{11}{\watt}};
    \node[anchor=west] at (cpu_max) {\fontsize{7}{12}\selectfont \SI{93}{\watt}};
    \node[anchor=west] at (lpFPGA_max) {\fontsize{7}{12}\selectfont \SI{0.1}{\watt}};

\end{tikzpicture}
}
	\caption{Comparison of the power consumption of the \ac{cnn}-based equalizer, running on \ac{gpu}, \ac{cpu} and \ac{fpga}. PT refers to the PyTorch and TRT refers to the TensorRT models.}
	\label{fig:gpu_cpu_power}
\end{figure}

Besides throughput and latency, we compare the power of the different implementation approaches. For the \acp{fpga}, the power is given by the Vivado power estimation tool, whereas for the Nvidia RTX \ac{gpu} and the \ac{cpu} we measure the power using the PyJoules library \cite{Pyjoules2019}. For the Nvidia AGX platform, the power is obtained based on the jetson-stats package \cite{Jetsonstats}. The results are shown in Fig. \ref{fig:gpu_cpu_power}. It can be seen that the power consumption of the \ac{lp} \ac{fpga} solution is orders of magnitude lower as compared to all other approaches. The AGX platform's power consumption lies in between the two \ac{fpga} approaches. Its power consumption is nearly constant up to $10^6$ \ac{spb}, and from there on it starts to increase significantly. This is probably caused by under-utilization of the \ac{gpu} for the small \ac{cnn} topology for low batch sizes. The power consumption of the \ac{ht} \ac{fpga} is around $2 \times$ higher than that of the Nvidia AGX. For the high performance \ac{cpu} and \ac{gpu}, the power increases up to \SI{93}{\watt} and \SI{250}{\watt} respectively.

\subsubsection{Summary}

In summary, the comparison shows that our flexible \ac{fpga} architecture is able to provide efficient solutions for low-power and high-throughput application scenarios. In particular, the \ac{ht} \ac{fpga} achieves higher throughput than all other platforms, including the high-performance \ac{gpu}, even for very high \ac{spb}. Moreover, the \ac{lp} \ac{fpga} provides much lower power consumption as the other platforms. Main reason for the superior performance of the \acp{fpga} is probably the efficient \ac{cnn} inference hardware architecture, which is perfectly adapted to the specific \ac{cnn} topology. In contrast, the other platforms provide a more general architecture that can also be utilized for other algorithms and use cases. 

To conclude, the results show that the \ac{fpga} as a platform, in combination with a hardware architecture based on optimizations across all design layers, can provide a promising solution for implementing \ac{ann}-based algorithms for communications. 
}

\section{Conclusion}
\label{sec:conclusion}

In this work, we present the \ac{fpga} implementation of a high-throughput \ac{cnn}-based equalizer for optical communications. The implementation is based on optimization across all design layers, beginning with an extensive design-space exploration of the \ac{cnn}, followed by a detailed quantization analysis, and culminating in the design of an efficient hardware architecture. As a result, the high-throughput \ac{fpga} implementation of our equalizer achieves a \ac{ber} around $4\times$ lower than that of a conventional linear equalizer while meeting the high-throughput requirements of a \SI{40}{\giga Bd} communication channel. Moreover, we demonstrate the flexibility of our custom hardware architecture by successfully applying our approach to a magnetic recording with a focus on low cost and low power. 
Further, we present a framework that optimizes the sequence length per instance to reduce the equalizer's latency under given throughput constraints. Finally, we compare our hardware implementation to optimized \ac{cpu} and \ac{gpu} implementations and show that the \ac{ht} \ac{fpga} achieves a throughput that is three orders of magnitude higher than that of the high-performance \ac{gpu} for a similar batch size. Moreover, we demonstrate that the same hardware architecture can also be applied to a low-power scenario, where an embedded \ac{gpu} is outperformed in terms of power consumption.



\section*{Declarations}

\subsection*{Funding}
This work was carried out in the framework of the CELTIC-NEXT project AI-NET-ANTILLAS (C2019/3-3) and was funded by the German Federal Ministry of Education and Research (BMBF) under grant agreements 16KIS1316 and 16KIS1317 as well as under grant 16KISK004 (Open6GHuB).

\subsection*{Conflict of interest}
The authors have no competing interests as defined by Springer, or other interests that
might be perceived to influence the results and/or discussion reported in this paper.

\subsection*{Author contribution}
The authors' contributions are as follows:\\ 
JN: design space exploration, hardware implementation, result evaluation, wrote the original draft. CF, VL, SR: assembled the experimental setup, performed the experiments to generate the dataset. CF, VL, LS, SR, NW: detailed draft review and editing. SR, LS, NW: idea and formulation of research goals, acquisition of funding.

\bibliography{IEEEabrv,bib_bibtex}


\begin{thebibliography}{23}
\ifx \bisbn   \undefined \def \bisbn  #1{ISBN #1}\fi
\ifx \binits  \undefined \def \binits#1{#1}\fi
\ifx \bauthor  \undefined \def \bauthor#1{#1}\fi
\ifx \batitle  \undefined \def \batitle#1{#1}\fi
\ifx \bjtitle  \undefined \def \bjtitle#1{#1}\fi
\ifx \bvolume  \undefined \def \bvolume#1{\textbf{#1}}\fi
\ifx \byear  \undefined \def \byear#1{#1}\fi
\ifx \bissue  \undefined \def \bissue#1{#1}\fi
\ifx \bfpage  \undefined \def \bfpage#1{#1}\fi
\ifx \blpage  \undefined \def \blpage #1{#1}\fi
\ifx \burl  \undefined \def \burl#1{\textsf{#1}}\fi
\ifx \doiurl  \undefined \def \doiurl#1{\url{https://doi.org/#1}}\fi
\ifx \betal  \undefined \def \betal{\textit{et al.}}\fi
\ifx \binstitute  \undefined \def \binstitute#1{#1}\fi
\ifx \binstitutionaled  \undefined \def \binstitutionaled#1{#1}\fi
\ifx \bctitle  \undefined \def \bctitle#1{#1}\fi
\ifx \beditor  \undefined \def \beditor#1{#1}\fi
\ifx \bpublisher  \undefined \def \bpublisher#1{#1}\fi
\ifx \bbtitle  \undefined \def \bbtitle#1{#1}\fi
\ifx \bedition  \undefined \def \bedition#1{#1}\fi
\ifx \bseriesno  \undefined \def \bseriesno#1{#1}\fi
\ifx \blocation  \undefined \def \blocation#1{#1}\fi
\ifx \bsertitle  \undefined \def \bsertitle#1{#1}\fi
\ifx \bsnm \undefined \def \bsnm#1{#1}\fi
\ifx \bsuffix \undefined \def \bsuffix#1{#1}\fi
\ifx \bparticle \undefined \def \bparticle#1{#1}\fi
\ifx \barticle \undefined \def \barticle#1{#1}\fi
\bibcommenthead
\ifx \bconfdate \undefined \def \bconfdate #1{#1}\fi
\ifx \botherref \undefined \def \botherref #1{#1}\fi
\ifx \url \undefined \def \url#1{\textsf{#1}}\fi
\ifx \bchapter \undefined \def \bchapter#1{#1}\fi
\ifx \bbook \undefined \def \bbook#1{#1}\fi
\ifx \bcomment \undefined \def \bcomment#1{#1}\fi
\ifx \oauthor \undefined \def \oauthor#1{#1}\fi
\ifx \citeauthoryear \undefined \def \citeauthoryear#1{#1}\fi
\ifx \endbibitem  \undefined \def \endbibitem {}\fi
\ifx \bconflocation  \undefined \def \bconflocation#1{#1}\fi
\ifx \arxivurl  \undefined \def \arxivurl#1{\textsf{#1}}\fi
\csname PreBibitemsHook\endcsname

\bibitem[\protect\citeauthoryear{Bayvel et~al.}{2016}]{Bayvel2016}
\begin{barticle}
\bauthor{\bsnm{Bayvel}, \binits{P.}},
\bauthor{\bsnm{Maher}, \binits{R.}},
\bauthor{\bsnm{Xu}, \binits{T.}},
\bauthor{\bsnm{Liga}, \binits{G.}},
\bauthor{\bsnm{Shevchenko}, \binits{N.A.}},
\bauthor{\bsnm{Lavery}, \binits{D.}},
\bauthor{\bsnm{Alvarado}, \binits{A.}},
\bauthor{\bsnm{Killey}, \binits{R.I.}}:
\batitle{Maximizing the optical network capacity}.
\bjtitle{Philosophical Transactions of the Royal Society A: Mathematical, Physical and Engineering Sciences}
\bvolume{374}(\bissue{2062}),
\bfpage{20140440}
(\byear{2016})
\end{barticle}
\endbibitem

\bibitem[\protect\citeauthoryear{Ahmet~Yazar}{2020}]{Yazar2020}
\begin{barticle}
\bauthor{\bsnm{Ahmet~Yazar}, \binits{H.A.} \bsuffix{Seda Dogan~Tusha}}:
\batitle{{6G} vision: An ultra-flexible perspective}.
\bjtitle{ITU Journal on Future and Evolving Technologies}
\bvolume{1},
\bfpage{121}--\blpage{140}
(\byear{2020})
\doiurl{10.52953/IKVY9186}
\end{barticle}
\endbibitem

\bibitem[\protect\citeauthoryear{Viswanathan and Mogensen}{2020}]{Viswanathan2020}
\begin{barticle}
\bauthor{\bsnm{Viswanathan}, \binits{H.}},
\bauthor{\bsnm{Mogensen}, \binits{P.E.}}:
\batitle{Communications in the {6G} era}.
\bjtitle{IEEE Access}
\bvolume{8},
\bfpage{57063}--\blpage{57074}
(\byear{2020})
\doiurl{10.1109/ACCESS.2020.2981745}
\end{barticle}
\endbibitem

\bibitem[\protect\citeauthoryear{Zerguine et~al.}{2001}]{zerguine2001}
\begin{barticle}
\bauthor{\bsnm{Zerguine}, \binits{A.}},
\bauthor{\bsnm{Shafi}, \binits{A.}},
\bauthor{\bsnm{Bettayeb}, \binits{M.}}:
\batitle{Multilayer perceptron-based {DFE} with lattice structure}.
\bjtitle{{IEEE} Transactions on Neural Networks}
\bvolume{12}(\bissue{3}),
\bfpage{532}--\blpage{545}
(\byear{2001})
\end{barticle}
\endbibitem

\bibitem[\protect\citeauthoryear{Schaedler et~al.}{2019}]{schaedler2019}
\begin{botherref}
\oauthor{\bsnm{Schaedler}, \binits{M.}},
\oauthor{\bsnm{Bluemm}, \binits{C.}},
\oauthor{\bsnm{Kuschnerov}, \binits{M.}},
\oauthor{\bsnm{Pittalà}, \binits{F.}},
\oauthor{\bsnm{Calabrò}, \binits{S.}},
\oauthor{\bsnm{Pachnicke}, \binits{S.}}:
Deep neural network equalization for optical short reach communication.
Applied Sciences
\textbf{9}(21)
(2019)
\end{botherref}
\endbibitem

\bibitem[\protect\citeauthoryear{Ney et~al.}{2022}]{ney2022}
\begin{botherref}
\oauthor{\bsnm{Ney}, \binits{J.}},
\oauthor{\bsnm{Hammoud}, \binits{B.}},
\oauthor{\bsnm{Dörner}, \binits{S.}},
\oauthor{\bsnm{Herrmann}, \binits{M.}},
\oauthor{\bsnm{Clausius}, \binits{J.}},
\oauthor{\bsnm{{ten Brink}}, \binits{S.}},
\oauthor{\bsnm{Wehn}, \binits{N.}}:
Efficient {FPGA} implementation of an {ANN}-based demapper using cross-layer analysis.
Electronics
\textbf{11}(7)
(2022)
\end{botherref}
\endbibitem

\bibitem[\protect\citeauthoryear{Khan et~al.}{2019}]{Khan2019}
\begin{barticle}
\bauthor{\bsnm{Khan}, \binits{F.N.}},
\bauthor{\bsnm{Fan}, \binits{Q.}},
\bauthor{\bsnm{Lu}, \binits{C.}},
\bauthor{\bsnm{Lau}, \binits{A.P.T.}}:
\batitle{An optical communication's perspective on machine learning and its applications}.
\bjtitle{Journal of Lightwave Technology}
\bvolume{37}(\bissue{2}),
\bfpage{493}--\blpage{516}
(\byear{2019})
\end{barticle}
\endbibitem

\bibitem[\protect\citeauthoryear{Amari et~al.}{2017}]{Amari2017}
\begin{barticle}
\bauthor{\bsnm{Amari}, \binits{A.}},
\bauthor{\bsnm{Dobre}, \binits{O.A.}},
\bauthor{\bsnm{Venkatesan}, \binits{R.}},
\bauthor{\bsnm{Kumar}, \binits{O.S.S.}},
\bauthor{\bsnm{Ciblat}, \binits{P.}},
\bauthor{\bsnm{Jaouën}, \binits{Y.}}:
\batitle{A survey on fiber nonlinearity compensation for 400 {Gb/s} and beyond optical communication systems}.
\bjtitle{IEEE Communications Surveys \& Tutorials}
\bvolume{19}(\bissue{4}),
\bfpage{3097}--\blpage{3113}
(\byear{2017})
\doiurl{10.1109/COMST.2017.2719958}
\end{barticle}
\endbibitem

\bibitem[\protect\citeauthoryear{Kuon and Rose}{2010}]{kuon2010quantifying}
\begin{bbook}
\bauthor{\bsnm{Kuon}, \binits{I.}},
\bauthor{\bsnm{Rose}, \binits{J.}}:
\bbtitle{Quantifying and Exploring the Gap Between FPGAs and ASICs}.
\bpublisher{Springer},
\blocation{New York}
(\byear{2010})
\end{bbook}
\endbibitem

\bibitem[\protect\citeauthoryear{Freire et~al.}{2023}]{Freire2022}
\begin{barticle}
\bauthor{\bsnm{Freire}, \binits{P.J.}},
\bauthor{\bsnm{Srivallapanondh}, \binits{S.}},
\bauthor{\bsnm{Anderson}, \binits{M.}},
\bauthor{\bsnm{Spinnler}, \binits{B.}},
\bauthor{\bsnm{Bex}, \binits{T.}},
\bauthor{\bsnm{Eriksson}, \binits{T.A.}},
\bauthor{\bsnm{Napoli}, \binits{A.}},
\bauthor{\bsnm{Schairer}, \binits{W.}},
\bauthor{\bsnm{Costa}, \binits{N.}},
\bauthor{\bsnm{Blott}, \binits{M.}},
\bauthor{\bsnm{Turitsyn}, \binits{S.K.}},
\bauthor{\bsnm{Prilepsky}, \binits{J.E.}}:
\batitle{Implementing neural network-based equalizers in a coherent optical transmission system using field-programmable gate arrays}.
\bjtitle{Journal of Lightwave Technology}
\bvolume{41}(\bissue{12}),
\bfpage{3797}--\blpage{3815}
(\byear{2023})
\doiurl{10.1109/JLT.2023.3272011}
\end{barticle}
\endbibitem

\bibitem[\protect\citeauthoryear{Kaneda et~al.}{2022}]{Kaneda2022}
\begin{barticle}
\bauthor{\bsnm{Kaneda}, \binits{N.}},
\bauthor{\bsnm{Chuang}, \binits{C.-Y.}},
\bauthor{\bsnm{Zhu}, \binits{Z.}},
\bauthor{\bsnm{Mahadevan}, \binits{A.}},
\bauthor{\bsnm{Farah}, \binits{B.}},
\bauthor{\bsnm{Bergman}, \binits{K.}},
\bauthor{\bsnm{Van~Veen}, \binits{D.}},
\bauthor{\bsnm{Houtsma}, \binits{V.}}:
\batitle{Fixed-point analysis and {FPGA} implementation of deep neural network based equalizers for high-speed {PON}}.
\bjtitle{Journal of Lightwave Technology}
\bvolume{40}(\bissue{7}),
\bfpage{1972}--\blpage{1980}
(\byear{2022})
\end{barticle}
\endbibitem

\bibitem[\protect\citeauthoryear{Li et~al.}{2021}]{Li2021}
\begin{barticle}
\bauthor{\bsnm{Li}, \binits{M.}},
\bauthor{\bsnm{Zhang}, \binits{W.}},
\bauthor{\bsnm{Chen}, \binits{Q.}},
\bauthor{\bsnm{He}, \binits{Z.}}:
\batitle{High-throughput hardware deployment of pruned neural network based nonlinear equalization for {100-Gbps} short-reach optical interconnect}.
\bjtitle{Opt. Lett.}
\bvolume{46}(\bissue{19}),
\bfpage{4980}--\blpage{4983}
(\byear{2021})
\end{barticle}
\endbibitem

\bibitem[\protect\citeauthoryear{Huang et~al.}{2022}]{Huang2022}
\begin{barticle}
\bauthor{\bsnm{Huang}, \binits{X.}},
\bauthor{\bsnm{Zhang}, \binits{D.}},
\bauthor{\bsnm{Hu}, \binits{X.}},
\bauthor{\bsnm{Ye}, \binits{C.}},
\bauthor{\bsnm{Zhang}, \binits{K.}}:
\batitle{Low-complexity recurrent neural network based equalizer with embedded parallelization for {100-Gbit/s/}\textlambda {PON}}.
\bjtitle{Journal of Lightwave Technology}
\bvolume{40}(\bissue{5}),
\bfpage{1353}--\blpage{1359}
(\byear{2022})
\end{barticle}
\endbibitem

\bibitem[\protect\citeauthoryear{Ney et~al.}{2023a}]{Ney2023}
\begin{bchapter}
\bauthor{\bsnm{Ney}, \binits{J.}},
\bauthor{\bsnm{Lauinger}, \binits{V.}},
\bauthor{\bsnm{Schmalen}, \binits{L.}},
\bauthor{\bsnm{Wehn}, \binits{N.}}:
\bctitle{Unsupervised {ANN}-based equalizer and its trainable {FPGA} implementation}.
In: \bbtitle{2023 Joint European Conference on Networks and Communications \& 6G Summit (EuCNC/6G Summit)},
pp. \bfpage{60}--\blpage{65}
(\byear{2023}).
\doiurl{10.1109/EuCNC/6GSummit58263.2023.10188269}
\end{bchapter}
\endbibitem

\bibitem[\protect\citeauthoryear{Ney et~al.}{2023b}]{Ney2023_2}
\begin{bchapter}
\bauthor{\bsnm{Ney}, \binits{J.}},
\bauthor{\bsnm{F{\"u}llner}, \binits{C.}},
\bauthor{\bsnm{Lauinger}, \binits{V.}},
\bauthor{\bsnm{Schmalen}, \binits{L.}},
\bauthor{\bsnm{Randel}, \binits{S.}},
\bauthor{\bsnm{Wehn}, \binits{N.}}:
\bctitle{From algorithm to implementation: Enabling high-throughput {CNN}-based equalization on {FPGA} for optical communications}.
In: \beditor{\bsnm{Silvano}, \binits{C.}},
\beditor{\bsnm{Pilato}, \binits{C.}},
\beditor{\bsnm{Reichenbach}, \binits{M.}} (eds.)
\bbtitle{Embedded Computer Systems: Architectures, Modeling, and Simulation},
pp. \bfpage{158}--\blpage{173}.
\bpublisher{Springer},
\blocation{Cham}
(\byear{2023})
\end{bchapter}
\endbibitem

\bibitem[\protect\citeauthoryear{Owaki and Nakamura}{2016}]{Owaki2016}
\begin{bchapter}
\bauthor{\bsnm{Owaki}, \binits{S.}},
\bauthor{\bsnm{Nakamura}, \binits{M.}}:
\bctitle{Equalization of optical nonlinear waveform distortion using neural-network based digital signal processing}.
In: \bbtitle{2016 21st OptoElectronics and Communications Conference (OECC) Held Jointly with 2016 International Conference on Photonics in Switching (PS)},
pp. \bfpage{1}--\blpage{3}
(\byear{2016})
\end{bchapter}
\endbibitem

\bibitem[\protect\citeauthoryear{Estaran et~al.}{2016}]{Estaran2016}
\begin{bchapter}
\bauthor{\bsnm{Estaran}, \binits{J.}},
\bauthor{\bsnm{Rios-Mueller}, \binits{R.}},
\bauthor{\bsnm{Mestre}, \binits{M.A.}},
\bauthor{\bsnm{Jorge}, \binits{F.}},
\bauthor{\bsnm{Mardoyan}, \binits{H.}},
\bauthor{\bsnm{Konczykowska}, \binits{A.}},
\bauthor{\bsnm{Dupuy}, \binits{J.-Y.}},
\bauthor{\bsnm{Bigo}, \binits{S.}}:
\bctitle{Artificial neural networks for linear and non-linear impairment mitigation in high-baudrate {IM/DD} systems}.
In: \bbtitle{ECOC 2016; 42nd European Conference on Optical Communication},
pp. \bfpage{1}--\blpage{3}
(\byear{2016})
\end{bchapter}
\endbibitem

\bibitem[\protect\citeauthoryear{Eriksson et~al.}{2017}]{Eriksson2017}
\begin{barticle}
\bauthor{\bsnm{Eriksson}, \binits{T.A.}},
\bauthor{\bsnm{Bülow}, \binits{H.}},
\bauthor{\bsnm{Leven}, \binits{A.}}:
\batitle{Applying neural networks in optical communication systems: Possible pitfalls}.
\bjtitle{IEEE Photonics Technology Letters}
\bvolume{29}(\bissue{23}),
\bfpage{2091}--\blpage{2094}
(\byear{2017})
\doiurl{10.1109/LPT.2017.2755663}
\end{barticle}
\endbibitem

\bibitem[\protect\citeauthoryear{Proakis and Salehi}{2008}]{proakis2008digital}
\begin{bbook}
\bauthor{\bsnm{Proakis}, \binits{J.G.}},
\bauthor{\bsnm{Salehi}, \binits{M.}}:
\bbtitle{Digital Communications, 5th Edition}.
\bpublisher{McGraw-Hill Higher Education},
\blocation{New York}
(\byear{2008})
\end{bbook}
\endbibitem

\bibitem[\protect\citeauthoryear{Nikolic et~al.}{2020}]{nikolic2020}
\begin{botherref}
\oauthor{\bsnm{Nikolic}, \binits{M.}},
\oauthor{\bsnm{Hacene}, \binits{G.B.}},
\oauthor{\bsnm{Bannon}, \binits{C.}},
\oauthor{\bsnm{Lascorz}, \binits{A.D.}},
\oauthor{\bsnm{Courbariaux}, \binits{M.}},
\oauthor{\bsnm{Bengio}, \binits{Y.}},
\oauthor{\bsnm{Gripon}, \binits{V.}},
\oauthor{\bsnm{Moshovos}, \binits{A.}}:
Bitpruning: Learning bitlengths for aggressive and accurate quantization.
arXiv Prepint
(2020)
\end{botherref}
\endbibitem

\bibitem[\protect\citeauthoryear{Araujo et~al.}{2019}]{Araujo2019}
\begin{botherref}
\oauthor{\bsnm{Araujo}, \binits{A.}},
\oauthor{\bsnm{Norris}, \binits{W.D.}},
\oauthor{\bsnm{Sim}, \binits{J.}}:
Computing receptive fields of convolutional neural networks.
Distill
(2019)
\end{botherref}
\endbibitem

\bibitem[\protect\citeauthoryear{Belgaid et~al.}{2019}]{Pyjoules2019}
\begin{botherref}
\oauthor{\bsnm{Belgaid}, \binits{M.c.}},
\oauthor{\bsnm{Rouvoy}, \binits{R.}},
\oauthor{\bsnm{Seinturier}, \binits{L.}}:
{Pyjoules: Python library that measures python code snippets}
(2019).
\url{https://github.com/powerapi-ng/pyJoules}
\end{botherref}
\endbibitem

\bibitem[\protect\citeauthoryear{Bonghi}{}]{Jetsonstats}
\begin{botherref}
\oauthor{\bsnm{Bonghi}, \binits{R.}}:
jetson-stats.
\url{https://github.com/rbonghi/jetson_stats}
\end{botherref}
\endbibitem

\end{thebibliography}

\end{document}